\documentclass[twocolumn]{aastex7}
\usepackage{amsmath}
\usepackage{longtable}
\newcommand{\minus}{\scalebox{1.0}{\bf{$-$}}}
\newcommand{\plus}{\scalebox{1.1}{+}}
\newcommand{\ie}{\it i.e.,}
\newcommand{\eg}{\it e.g.,}

\begin{document}
\title{Intra-night optical variability and radio characteristics of extremely radio-loud narrow-line Seyfert 1 galaxies}

\author[orcid=0000-0002-6040-4993,sname='Singh']{Veeresh Singh}
\affiliation{Astronomy and Astrophysics Division, Physical Research Laboratory, Navrangpura, Ahmedabad, Gujarat-380 009, India}
\email[show]{veeresh@prl.res.in}

\author[orcid=0000-0001-5878-4811, sname='Kumar']{Parveen Kumar}
\affiliation{Astronomy and Astrophysics Division, Physical Research Laboratory, Navrangpura, Ahmedabad, Gujarat-380 009, India}
\email[show]{physicssehrawat@gmail.com}

\author[orcid=0000-0002-9526-0870, sname='Das']{Avik Kumar Das}
\affiliation{Astronomy and Astrophysics Division, Physical Research Laboratory, Navrangpura, Ahmedabad, Gujarat-380 009, India}
\email{avikdas@prl.res.in}

\author[orcid=0000-0001-5878-4811, sname='Ojha']{Vineet Ojha}
\altaffiliation{Current affiliation : Kavli Institute for Astronomy and Astrophysics, Peking University, Beijing 100871, China}
\affiliation{Astronomy and Astrophysics Division, Physical Research Laboratory, Navrangpura, Ahmedabad, Gujarat-380 009, India}
\email{vineetojhabhu@gmail.com }

%
%
%
%
%

\begin{abstract}
Narrow-line Seyfert 1 galaxies (NLS1s) are generally known to be radio-quiet Active Galactic Nuclei (AGN), but
a tiny subset of them are found to be extremely radio-loud with radio loudness parameter ($R_{\rm 1.4~GHz}$) $>$ 100.
Given their rarity we investigated intra-night optical variability (INOV) and radio characteristics of a sample
of 16 extremely radio-loud NLS1s. For all but four sample sources we report intra-night photometric
monitoring for the first time with at least one monitoring session per source lasting for a minimum of
3.0 hours duration. In our sample, we detect INOV with a high duty cycle (up to 25 per cent) and large
average amplitude ($\overline{\psi}$ $\sim$ 0.16) similar to that found in blazars.
Using 3.0 GHz Very Large Array Sky Survey (VLASS) and auxiliary multi-frequency radio data we find that our RL-NLS1s
are luminous ($L_{\rm 3.0~GHz}$ $\geq$ 10$^{24}$ W~Hz$^{-1}$), compact (less than a few kpc), variable,
flat spectrum (${\alpha}_{\rm radio}$ $>$ -0.5) radio sources. The INOV, radio characteristics, and radio luminosity
($L_{\rm 1.4~GHz}$) versus super-massive black hole mass ($M_{\rm SMBH}$) plot infer that
extremely radio-loud NLS1s are low-$z$ and low-luminosity analogs of flat spectrum radio quasars wherein the
former are powered by, on average, one order-of-magnitude less massive SMBHs.
\end{abstract}



\section{Introduction}\label{sec:intro}
Narrow-line Seyfert 1 galaxies (NLS1s), a subclass of Active Galactic Nuclei (AGN), are characterised with the
relatively narrow width of broad permitted Balmer emission lines (FWHM(H$\beta$) $<$ 2000 km $s^{-1}$) in their optical spectra,
and flux ratio of [O III]$\lambda$5007$\AA$ to H$\beta$ line ([O III]/H$\beta$) $<$ 3~\citep{Pogge2000}.
NLS1s are mostly radio-quiet (RQ; $R$ $<$ 10) except a small fraction of nearly 7.0 per cent of them being
radio-loud (RL; $R$ $\ge$ 10)~\citep{Komossa06}; where radio-loudness parameter ($R$) is defined as the ratio
of 5.0 GHz radio flux density to the optical continuum flux at 4400{\AA}.
The fraction of extreme RL-NLS1s with $R$ $>$ 100 is even lower {\ie} merely 2.5 per cent \citep{Singh18}.
The scarcity of RL-NLS1s, in particular extremely radio-loud RL-NLS1s, is intriguing.
Unlike conventional RL-AGN, RL-NLS1s seem to be powered
by SMBHs of relatively lower masses (10$^6$ $-$ 10$^{8}$~M$\odot$) and show higher Eddington ratios
($>$ 0.05) \citep{Deo06,Peterson11,Wang14}. Although, possibility of systematic underestimation of the mass of SMBH
in NLS1s is not completely ruled out \citep{Marconi08,Calderone13,Ojha20b}.
The existence of relativistic jets in RL-NLS1s with SMBHs of relatively lower masses poses
a challenge to the conventional theoretical paradigm of jet formation according to which only SMBHs with
mass $>$ 10$^8$~M$\odot$ are capable of launching relativistic jets \citep{Bottcher02}.
Thus, RL-NLS1s with extreme radio-loudness ($R$ $>$ 100) warrant investigation to understand their nature.
\par
To gain insights into the nature of RL-NLS1s there have been attempts to directly image parsec-scale radio-jets in them using
high-resolution radio observations such as those from Very Long Baseline Array (VLBA\footnote{https://public.nrao.edu/telescopes/vlba/})
\citep{Doi11,Gu15}. The VLBA observations detecting one-sided jet emerging from the compact core confirm
the presence of relativistic jet in at least some RL-NLS1s \citep{Shao23}.
In fact, RL-NLS1s are known to exhibit diverse radio properties in terms of their radio structures and spectra.
For instance, using 5.0~GHz VLBA observations \cite{Gu15} detected core-jet structure in
only half of their sample of 14 extremely radio-loud ($R$ $>$ 100) NLS1s.
It is suggested that only a fraction their RL-NLS1s may possess relativistically beamed jets, while some of their sources
resembling with compact steep spectrum (CSS) radio sources may host only slow moving mildly relativistic jets.
The detection of $\gamma$-ray emission in a handful of NLS1s from the Fermi-Large Area Telescope (LAT\footnote{https://fermi.gsfc.nasa.gov/})
has also provided us clear evidence for the existence of relativistic jets in at least a small subset of NLS1s \citep{Abdo09a}.
All the $\gamma$-ray detected NLS1s are found to be radio-loud and variable \citep{Paliya19,Ojha21},
although, vice-versa is not true possibly due to the limited sensitivity of Fermi-LAT.
In fact, despite concerted efforts the number of $\gamma$-ray detected NLS1s have been limited only to nearly
two dozen of NLS1s \citep[e.g,][]{Abdo09a,DAmmando12,DAmmando15,Yao15,Yang18,Yao19,Paliya19a,Rakshit21,Li23}.
\par
Considering the limited detections of RL-NLS1s with VLBA and Fermi-LAT we exploit optical monitoring
as a complementary tool to infer the presence of relativistic jets in them. We also examine their
radio characteristics to gain further insights onto the nature of jets in them.
It is well known that, due to beaming effect, AGN with relativistic jets aligned close to the line-of-sight
exhibit flux variability on short timescales in the range of a few minutes to a few hours, known
as micro-variability \citep{Miller89} or intra-night optical variability (INOV)~\citep{Wagner-Witzel95,Ojha22}.
The jet-dominated AGN such as blazars are conventionally known to exhibit strong and rapid INOV characterised
with high amplitude and high duty-cycle \citep{Goyal13}. Hence, INOV studies have been used as a tool to
indicate the presence or absence of relativistic jets in different sub-classes of AGN {\eg}
quasars \citep{Jang97,Kumar15,Kumar16}, $\gamma$-ray detected NLS1s \citep{Paliya13,Ojha21},
RL-NLS1s \citep{Ojha22} and radio-quiet NLS1s \citep{Miller2000,Ojha22a,Gopal23}. In our study, we treat
INOV parameters ({\ie} amplitude and duty cycle) as robust indicators for the absence/presence of
relativistic jet in RL-NLS1s.
\par
The structure of this paper is as follows. In Section~\ref{sec:sample}, we describe our sample and its
basic properties. In Section~\ref{sec:observations}, we provide the details of our monitoring observations
and data reduction. In Section~\ref{sec:statistical}, we describe the statistical tests that are employed
to determine the presence of INOV in differential light curves (DLCs). Section~\ref{sec:INOV} provides a comparison
of RL-NLS1s INOV properties to that of the different samples of jetted AGN reported in the literature.
In Section~\ref{sec:radio}, we report the radio characteristics of our RL-NLS1s. Section~\ref{sec:Discussion}
is devoted to the discussion on inferences derived from the INOV and radio properties of RL-NLS1s.
In Section~\ref{sec:conclusions}, we present the conclusions of our work.
\\
In our study, we adopt cosmological parameters $H_{0}$ = 70~km~s$^{-1}$~Mpc$^{-1}$, ${\Omega}_{\rm M}$ = 0.27, ${\Omega}_{\Lambda}$ = 0.73.
The radio spectral index $\alpha$ is defined as S$_{\nu}$ $\propto$ ${\nu}^{\alpha}$, where S$_{\nu}$ is
the flux spectral density.

\section{The sample}
\label{sec:sample}
Our sample consists of 16 RL-NLS1s that are gleaned from a large sample of radio-detected NLS1s
reported in \cite{Singh18} who used 1.4 GHz Faint Images of the Radio Sky at Twenty-cm \citep[FIRST;][]{Becker95}
and 1.4 GHz NRAO VLA Sky Survey \citep[NVSS;][]{Condon98} observations to study radio properties of NLS1s.
Our RL-NLS1s satisfy following criteria : (i) they are classified as bonafide NLS1s based on their
Sloan Digital Sky Survey (SDSS) optical spectra \citep{Rakshit17}, (ii) they are extremely radio-loud with $R_{\rm 1.4~GHz}$ $>$ 100,
(iii) their $r$ band magnitudes ($m_{r}$) are $<$ 18.6 mag which makes
them suitable for monitoring with 1.0$-$2.0m class optical telescopes,
(iv) to maintain the uniformity in term of radio sizes,
we also excluded sources, if any, with large-scale double-lobe radio structures. None of our sources show
any extended radio emission in 1.4 GHz FIRST, 150 MHz TIFR GMRT Sky Survey \citep[TGSS;][]{Intema17} and 144 MHz LOFAR
Two-metre Sky Survey \citep[LoTSS;][]{Shimwell17} images, whenever available.
We note that, with these selection criteria, our sample is no more a representative of general population of NLS1s.
In fact, it is a subset of NLS1s with extreme properties.
\par
In Table~\ref{tab:sample} we list our sample sources and their basic parameters {\ie} source name based on
their RA (J2000) and DEC (J2000), $r$ band magnitude ($m_{r}$), redshift ($z$), 1.4 GHz flux
density ($S_{\rm 1.4~GHz}$), 1.4 GHz radio luminosity ($L_{\rm 1.4~GHz}$), monochromatic optical
luminosity at 5100{\AA} ($L_{\rm 5100{\AA}}$), radio loudness parameter ($R_{\rm 1.4~GHz}$), and
the mass of SMBH. The 5100{\AA} optical continuum luminosity is obtained by modelling the SDSS
spectrum \citep[see][]{Rakshit17}.
The radio-loudness parameter is estimated as the ratio of 1.4 GHz radio luminosity to 4400{\AA} optical luminosity,
($R_{\rm 1.4~GHz}$ = $\frac{{\nu}L_{\rm 1.4~GHz}}{{\nu}L_{\rm 4400~{\AA}}}$;
\citet{Singh18}). The monochromatic optical continuum luminosity at 4400{\AA} ($L_{{\lambda}4400}$) is interpolated
from 5100{\AA} luminosity ($L_{{\lambda}5100}$) assuming a power law optical spectrum with an average spectral
index of -0.5 \citep{Sikora07}. To estimate the radio-loudness parameter we used 1.4 GHz luminosity instead
of 5.0 GHz luminosity as radio observations at latter frequency are unavailable for the most of our sources.
All of our sample sources are extremely radio-loud with $R_{\rm 1.4~GHz}$ in the range of 106.6 to 250593.
The radio-loud character is also vindicated from their high 1.4~GHz radio luminosities spanning across
3.55 $\times$ 10$^{23}$ W Hz$^{-1}$ to 1.38 $\times$ 10$^{27}$ W~Hz$^{-1}$.
We note that four of our sample sources (J084957.98$+$510829.08 aka SBS 0846+513,
J094857.31+002225.51 aka PMN J0948+0022, J130522.74+511640.26 and J150506.47+032630.82 aka PKS 1502+036) are also detected in $\gamma$-ray with Fermi/LAT \citep[see][]{DAmmando12,Abdo09b,Paliya13}.
The presence of $\gamma$-ray detected NLS1s in our sample is not surprising as they tend to be extremely
radio-loud \citep{Orienti15,Shao23}.
\par
To estimate the mass of SMBH in our RL-NLS1s we have used virial relation
$M_{\rm SMBH} = f (\frac{R_{\rm BLR}{\Delta}{v^2}}{G})$,
where ${\Delta}v$ is the full width at half maxima (FWHM) of broad component of H$\beta$ emission line,
$R_{\rm BLR}$ is the distance of BLR cloud from SMBH, $f$ is the scaling factor which depends on the geometry
and kinematics of BLR and is considered to be 3/4 for spherically distributed BLR clouds \citep{Kaspi2000}.
The distance of BLR is estimated using $R_{\rm BLR}$ $-$ $L_{\rm 5100~{\AA}}$ scaling relation which
can be expressed as log$R_{\rm BLR}$ = $K$ + $\alpha$$^{\star}$log($L_{5100~{\AA}}$/10$^{44}$);
where $R_{\rm BLR}$ is in the unit of light days and $L_{5100~{\AA}}$ is in the unit of erg~s$^{-1}$.
For our calculation we used $K$ = 1.527 and $\alpha$ = 0.533 from \cite{Bentz13}. The masses of
SMBHs ($M_{\rm SMBH}$) in our RL-NLS1s are found to be in the range of 2.0 $\times$ 10$^{6}$ M$\odot$ to
9.33 $\times$ 10$^{7}$ M$\odot$.
\begin{table*}
\caption{Our RL-NLS1s sample and the basic parameters}
\label{tab:sample}
\hskip-2.0cm
\begin{tabular}{lccccccc}
\hline \hline
Source & $m_{r}$ & Redshift ($z$) & log$L_{\rm 5100~{\AA}}$ & $S_{\rm 1.4~GHz}^{\rm int}$ & log$L_{\rm 1.4~GHz}$ & $R_{\rm 1.4~GHz}$ & log$M_{\rm SMBH}$ \\
name                   &  (mag) &      & (erg~s$^{-1}$)  & (mJy)        &  (W~Hz$^{-1}$)       &   & (M$\odot$)  \\
(1)                    &  (2)   &  (4)   & (5)          & (6)      &  (7)            &     (8)    &  (9)    \\  \hline
J080000.05$+$152326.09 &  18.57$\pm$0.01 & 0.27416$\pm$0.00003  & 43.70 & 6.74$\pm$0.14    & 24.17$\pm$0.02 & 115.8    & 7.13$\pm$0.03  \\
J083454.90$+$553421.13 &  17.19$\pm$0.01 & 0.24153$\pm$0.00004  & 43.33 & 8360.63$\pm$1.16 & 27.14$\pm$0.02 & 250593.6 & 6.97$\pm$0.04  \\
J084957.98$+$510829.08$^{\gamma}$ & 18.28$\pm$0.01 & 0.58404$\pm$0.00013 & 44.22 & 350.53$\pm$0.15 & 26.63$\pm$0.02 & 11367.8 & 7.38$\pm$0.09  \\
J090409.65$+$581527.45 &  16.63$\pm$0.01 & 0.14609$\pm$0.00005   & 43.09 & 6.37$\pm$0.12 & 23.55$\pm$0.02 & 106.6 & 6.30$\pm$0.16  \\
J092533.67$+$021342.59 &  18.08$\pm$0.01 & 0.22801$\pm$0.00003 & 43.13 & 6.28$\pm$0.15    & 23.96$\pm$0.03 & 261.4    & 6.35$\pm$0.06  \\
J093323.02$-$001051.62 &  18.56$\pm$0.01 & 0.79577$\pm$0.00025 & 44.65 & 105.58$\pm$0.15  & 26.42$\pm$0.02 & 2786.5   & 7.56$\pm$0.08  \\
J094857.31$+$002225.51$^{\gamma}$ &  18.43$\pm$0.01 & 0.58384$\pm$0.00012 & 44.71 & 111.46$\pm$0.15 & 26.14$\pm$0.02 & 1168.7 & 7.30$\pm$0.02  \\
J103346.39$+$233220.00 &  18.51$\pm$0.01 & 0.46995$\pm$0.00004 & 44.40 & 9.20$\pm$0.14    & 24.83$\pm$0.02 & 114.9    & 7.27$\pm$0.03  \\
J111439.00$+$324134.00 &  17.13$\pm$0.01 & 0.18757$\pm$0.00002 & 43.58 & 107.77$\pm$0.12  & 25.01$\pm$0.02 & 1018.6   & 7.08$\pm$0.01  \\
J111752.42$+$213619.30 &  17.54$\pm$0.01 & 0.16921$\pm$0.00002 & 42.81 & 13.08$\pm$0.14   & 23.99$\pm$0.02 & 577.8    & 6.31$\pm$0.15  \\
J122039.35$+$171820.82 &  18.16$\pm$0.01 & 0.44302$\pm$0.00006 & 44.56 & 87.55$\pm$0.15   & 25.75$\pm$0.02 & 654.3    & 7.58$\pm$0.02  \\
J130522.74$+$511640.26$^{\gamma}$ &  17.10$\pm$0.01 & 0.78499$\pm$0.00008 & 45.31 & 86.94$\pm$0.15   & 26.33$\pm$0.02 & 484.8    & 7.97$\pm$0.03  \\
J143244.91$+$301435.33 &  18.56$\pm$0.01 & 0.35456$\pm$0.00006 & 44.12 & 49.98$\pm$0.14   & 25.29$\pm$0.02 & 599.9    & 7.43$\pm$0.01  \\
J150506.47$+$032630.82$^{\gamma}$ &  18.22$\pm$0.01 & 0.40827$\pm$0.00009 & 44.00 & 380.49$\pm$0.15  & 26.31$\pm$0.02 & 8462.0 & 7.04$\pm$0.08  \\
J153102.48$+$435637.69 &  17.09$\pm$0.01 & 0.45230$\pm$0.00002 &  44.91 & 55.45$\pm$0.15   & 25.58$\pm$0.02 & 194.8    & 7.68$\pm$0.03  \\
J160048.75$+$165724.41 &  18.09$\pm$0.01 & 0.37251$\pm$0.00006 &  44.41 & 30.16$\pm$0.15   & 25.12$\pm$0.02 & 208.9    & 7.59$\pm$0.02  \\
\hline
\end{tabular}
{\bf Note}. Source name is based on RA (J2000) and DEC (J2000).
The $r$ band magnitude and spectroscopic redshifts are taken from the SDSS. 1.4 GHz flux density is taken from
the FIRST. 1.4 GHz radio luminosity is $K$ corrected.
$^{\gamma}$ : $\gamma$-ray detected NLS1s.
\end{table*}

\section{Observations and data reduction}
\label{sec:observations}
\subsection{Intra-night photometric monitoring observations}
We performed photometric monitoring observations of 16 RL-NLS1s in broad-band Bessel $R$ band filter
using 1.2m telescope located at the Mount Abu
Observatory \citep{Deshpande95} which is operated by Physical Research Laboratory, Ahmedabad, India.
Our sample sources are monitored during December 2019 $-$ April 2021 for at least one session lasting more than 3.0 hours,
using 1K $\times$ 1K ANDOR CCD detector
of Mount-Abu Faint Object Spectrograph and Camera $-$ Pathfinder \citep[MFOSC-P;][]{Srivastava21}, which is equipped
with Bessel's $B$, $V$, $R$, $I$ and narrow-band H${\alpha}$ filters, mounted at $f/13$ cassegrain focus of the telescope.
The CCD of MFOSC-P has pixel size of 13~$\mu$m, plate scale of 3.3 pixels per arcsec, and
field-of-view (FOV) of 5$^{\prime}$.2 $\times$ 5$^{\prime}$.2.
More recently in 2025, we monitored five of our sample sources with one session lasting more than 3.0 hour on each source
using 11K $\times$ 8.0K front-illuminated CMOS imager mounted on 1.2m telescope. The imager equipped
with $U$, $B$, $V$, $R$, $I$ filters provides 9$^{\prime}$.0 $\times$ 8$^{\prime}$.0 FoV.
The CMOS imager has 3.76 $\mu$m pixel size and gives a dark current of 0.003 e$^{-}$ pixel$^{-1}$ sec$^{-1}$ and a
readout noise of approximately 1e$^{-}$ at 0$^{\circ}$C.
We opted to perform photometric monitoring in $R$ band in which sensitivity of CCD as well as CMOS detector is maximum.
In Table~\ref{tab:log}, we provide observational log giving details on the dates and duration of observations, magnitudes and colors ($g-r$) of target sources and comparison stars.
\par
During observations we set exposure time for each science frame around $3.0$ to $8.0$ minutes depending
upon the source brightness and sky conditions. In all except two monitoring sessions
we obtained signal-to-noise ratio (SNR) of 25 or higher. The poor sky conditions compounded with somewhat less exposure time
of 3.0 minutes resulted SNR of 5.0$-$10 during 25th and 26th January 2025.
For each night we acquired sky flat-field images taken at the dusk as
well as dawn, and at least three bias frames. The dark frames were not required for our observations
due to relatively low temperature of the CCD detector {\ie} CCD is cooled to -80 degree $^{\circ}$C using
Peltier cooling. During our observations seeing (FWHM) is found to
be 1$^{\prime\prime}$.5 $-$ 3$^{\prime\prime}$.0 with an average value of nearly 2$^{\prime\prime}$.0.
\subsection{Data Reduction}
We reduced raw CCD images using the standard tasks available in the Image Reduction and Analysis Facility (IRAF\footnote{http://iraf.noao.edu/})
which include bias subtraction, flat-fielding and cosmic-ray removal. To determine the instrumental magnitudes of target source and comparison stars in each
CCD frame we performed aperture photometry~\citep{Stetson87,Stetson92} using the Dominion Astronomical Observatory
Photometry \textrm{II} (DAOPHOT II\footnote{http://www.astro.wisc.edu/sirtf/daophot2.pdf}) algorithm.
We note that the choice of aperture size is a crucial parameter in the aperture photometry as it decides the SNR
and the amount of contamination from host galaxy.
To choose the aperture size for photometry, we first determined the mean PSF
by averaging the FWHMs of five moderately bright unsaturated stars in each frame and then taking the mean of FWHMs of
all the frames.
In some of the previous studies it is shown that the aperture size taken approximately equal to the FWHM of PSF gives the highest
SNR \citep[see][]{Howell88}. However, such a small aperture size can also give rise spurious variability from the host galaxy, if AGN plus
host galaxy system is not an unresolved source and seeing varies significantly~\citep[see][]{Cellone2000}.
Therefore, to mitigate seeing-induced spurious variability we chose an optimum aperture size obtained by
performing aperture photometry with varying radii set equal to $n$ $\times$ FWHM (where $n$ is an integer and varies from 1 to 7).
As expected, we find an increase in flux/magnitude with increase in aperture size but flux/magnitude stabilizes
at the aperture size of a few times of FWHM. For most of our sample sources we found an optimum aperture size
(diameter) of 2 $\times$ FWHM, which is similar to previous works on NLS1s \citep[see][]{Ojha22}.
\par
It is important to note that several of our sample sources are at the moderate redshifts ($z$ $<$ 0.5) and their host galaxies
can have angular sizes more than the aperture size (two times of average FWHM) causing DLC to mimic FWHM (seeing) variation
due to varying contribution from the underlying host galaxy \citep{Cellone2000}.
Using deep imaging of NLS1s host galaxies \cite{Olguin-Iglesias20} showed that only
relatively high$-z$ sources with $z$ $>$ 0.5 can be regarded as safe against the seeing induced spurious variations.
Therefore, to avoid seeing induced spurious variations in case of low-$z$ sources ($z$ $<$ 0.5) that show extended hosts, we used
fixed aperture size equivalent to two times of average FWHM.
The inspection of SDSS optical images reveals that all except five sample sources (J083454.90+553421.13 at $z$ = 0.242,
J090409.65+581527.45 at $z$ = 0.146, J092533.67+021342.59 at $z$ = 0.228, J111439.00+324134.00 at $z$ = 0.187, and
J111752.42+213619.30 at $z$ = 0.169) show only strong nuclear component, similar to quasars,
and lack a prominent host galaxy component.
\par
To investigate variability in our RL-NLS1s we plot DLCs using instrumental magnitudes of
target source and comparison stars. Based on the two suitably chosen steady comparison stars (named as `S1' and `S2')
that are close to the target source (named as `AGN') in terms of position as well as magnitude we have two DLCs
(`AGN-S1' and `AGN-S2') for each monitoring session (see Figure~\ref{fig:DLCs}).
To demonstrate the non variable nature of comparison stars we also plot `S1-S2' DLC.
The details of chosen comparison stars  ({\ie} position, magnitudes, and `$g-r$' colors) are listed in
Table~\ref{tab:log}.
We note that, in all but three of our monitoring sessions we find at least one comparison star within $\pm$1.0 mag of the target
source. For remaining three sessions targeted on J093323.01-001051.6, J094857.31+002225.5, J103346.39+233220.0
we find a comparison star within $\pm$1.6 mag.
For all the sessions, $g-r$ color difference between the target RL-NLS1 and its comparison stars is also
within 1.5 mag. For color difference of this order,
the changing atmospheric attenuation during a session produces a negligible effect on the
DLCs \citep{Carini92,Stalin04,Ojha21}.

\section{Statistical tests for examining intra-night optical varaibilty}
\label{sec:statistical}
In the literature, various statistical tests have been proposed to examine the presence of INOV in AGN
{\eg} $C$-statistic \citep{Jang97}, one-way analysis of variance \citep[ANOVA;][]{Diego10}, and $F$-test \citep{Villforth10,Goyal12}. The different statistical tests have their limitations and advantages.
For instance, \citet{Diego10} argued that $C$-statistics does not have a Gaussian distribution and
the critical value of 2.576 used for confirming the presence of variability at $3\sigma$ level is too conservative.
On the other hand, ANOVA test requires a rather large number of data points in the DLC to
have several points within each sub-group used for the analysis.
The $F$-test requires data to follow Gaussian distribution. We performed Anderson-Darling test
\citep{Anderson52} and found that the Differential Light Curves (DLCs) of all except one
(J111752.42+213619.30) sample sources follow Gaussian distribution.
Considering the suitability of $F-$test for our data in which difference between target and comparison star brightness is
mostly within 1.0 mag we employ two versions of the $F$-test {\ie} standard version of $F$-test (hereafter $F^{\eta}-$test) and power-enhanced $F$-test (hereafter $F_{\rm enh}$-test). In the following subsections we provide a brief description of the two versions of $F-$test.
\par
In addition to the $F$-test we also apply Chi-square (${\chi}^2$)-test and Auto-Regressive Integrated Moving
Average (ARIMA) models (see section~\ref{sec:ARIMA}) to detect variability in DLCs. In the ${\chi}^2$-test,
we fit each DLC with a constant equal to their mean
value and estimate reduced ${\chi}^2$ and $p$-value of the null hypotheses {\ie} no varaibilty. The
reduce ${\chi}^2$ is defined as $\frac{1}{N-1}$~$\sum_{i=0}^{N}$ $\frac{(O_{\rm i} - {\mu})^2}{{\epsilon}_{\rm i}^2}$, where $O_{\i}$
is $i$-th observed value, $\mu$ is the mean value, ${\epsilon}_{\rm i}$ is error associated with an individual datum value and $N$ is
total number of data points in a DLC.

\subsection{The standard $F$-test ($F^{\eta}$ test)}
\label{F-testStd}
The $F^{\eta}-$test can be expressed as :
\begin{multline}
\label{eq.fetest}
F_{1}^{\eta} = \frac{\sigma^{2}_{\rm (AGN-s1)}} { \eta^2 \langle \sigma_{\rm AGN-s1,~err}^2 \rangle},
\hspace{0.2cm} F_{2}^{\eta} = \frac{\sigma^{2}_{\rm (AGN-s2)}} { \eta^2 \langle \sigma_{\rm AGN-s2,~err}^2 \rangle},\\
\\
F_{s1-s2}^{\eta} = \frac{\sigma^{2}_{\rm (s1-s2)}} { \eta^2 \langle \sigma_{\rm s1-s2,~err}^2 \rangle},
\end{multline}

\citep[see][]{Goyal12}
where ${\sigma}^{2}_{\rm (AGN-s1)}$, ${\sigma}^{2}_{\rm (AGN-s2)}$ and ${\sigma}^{2}_{\rm (S1-s2)}$  are the variances of `AGN$-$S1', `AGN$-$S2' and `S1-S2' DLCs,
and $\langle \sigma_{\rm AGN-s1,~err}^2 \rangle = \sum_{i=0}^{N}\frac{{\sigma}^{2}_{i,err}{\rm (AGN-s1)}}{N}$,
$\langle \sigma_{\rm AGN-s2,~err}^2 \rangle = \sum_{i=0}^{N}\frac{{\sigma}^{2}_{i,err}{\rm (q-s2)}}{N}$ and
$\langle \sigma_{\rm s1-s2,~err}^2 \rangle = \sum_{i=0}^{N}\frac{{\sigma}^{2}_{i,err}{\rm (s1-s2)}}{N}$
are mean square of errors on the individual data points in `AGN$-$S1', `AGN$-$S2' and `S1$-$S2' DLCs, respectively.
While applying the statistical tests, we account for the fact that the photometric errors on individual data points in a given DLC,
as returned by the IRAF and DAOPHOT software routines are systematically lower by a factor $\eta$
ranging between $1.3$ and $1.75$, as reported in various studies \citep[e.g.,][] {Stalin04b,Bachev05}.
We use the best-fit value $\eta$ = $1.54$ obtained from 262 intranight monitoring sessions of AGN in \cite{Goyal13}.
\par
To apply $F^{\eta}$-test we compute $F$-values for both `AGN$-$S1' and `AGN$-$S2' DLCs for each session
and compare it with the critical value $F_{c}$ = $F^{(\alpha)}_{\nu_{AGN-s}}$, where $\alpha$ is the significance
level set for the test, and $\nu_{AGN-s}$ is the
degree of freedom ($N$ $-$ 1) of the DLC. We opt for the two values of significance level $\alpha=$ 0.01 and 0.05
that correspond to confidence levels of greater than 99 and 95 per cent, respectively.
If the computed $F$ value exceeds the corresponding critical value $F_{c}$, the null hypothesis ({\ie} no variability)
is discarded to the respective level of confidence.
We consider target source as variable (`V') if the computed $F$-values for both of its DLCs
(`AGN-S1' and `AGN-S2') are $\ge F_{c}(0.99)$, which corresponds to $99$ per cent confidence level.
The target source is termed as non-variable (`NV'), if either of the two DLCs is found to have
an $F$-value $< F_{c}(0.95)$. The remaining cases are classified as probably variable (`PV').
In Table~\ref{tab:INOVTests}, we list $F$ values  ($F^{\eta}_{1}$ and $F^{\eta}_{2}$) corresponding to both the DLCs,
and the INOV status for all target sources in each monitoring session.
We also checked the presence of variability in the DLC of each set of comparison stars in each session
by computing $F_{S1-S2}^{\eta}$ value. The DLCs of each set of comparison stars in all sessions are found be
non-variable which ensures the reliability of INOV status of our target sources.
We also list averaged photometric error ($\sqrt{\sigma_{i,err}^2}$) in the two DLCs (`AGN-S1 and `AGN-S2') of
target sources, which is around 0.03 $-$ 0.04 mag for our sample sources.
\subsection{The enhanced $F$-test ($F_{\rm enh}$-test)}
\label{F-testEnh}
Unlike standard $F$-test, the enhanced $F_{\rm enh}$-test accounts for the difference in the brightness levels of target AGN
and comparison stars. It enhances the power of $F$-test by considering several comparison stars, and consequently it mitigates the
possibility of spurious variability introduced by a single comparison star which may not be completely steady.
The $F_{\rm enh}$-test can be expressed as
\begin{equation}
\label{eq:EnFtest}
F_{\rm enh} = \frac{{\sigma}^{2}_{\rm AGN - ref}}{{\sigma}^{2}_{\rm comb}}.
\end{equation}
\citep[see][]{deDiego14} where, ${\sigma}^{2}_{\rm AGN}$ is the variance in the `AGN$-$reference star’ DLC, and ${\sigma}^{2}_{\rm comb}$ is the combined
variance of `comparison star$-$reference star' DLCs.
The  ${\sigma}^{2}_{\rm comb}$ can be expressed as
\begin{equation}
\label{eq:EnFtestrms}
{\sigma}^{2}_{\rm comb} = \frac{1}{(\sum_{j=1}^{k}N_{j} -k)}{\sum_{j=1}^{k}}\sum_{i=1}^{N_j}{\omega}_{j} (m_{j,i} - \overline{m}_{j})^{2}
\end{equation}
where $N_{j}$ are number of data points in $j$th comparison star and $k$ is number of comparison stars,
$m_{j, i}$ is the $j^{th}$ `comparison star $-$ reference star' differential instrumental magnitudes
and $\overline{m}_{j}$ is average magnitude of the corresponding DLC with $N_{j}$ data points.
In our analysis we chose reference star from the two comparison stars {\ie} the one which is matching better
in magnitude to the target AGN, and therefore $k$ = 1.
The ${\omega}_{j}$ represents a scaling factor which accounts for the fact that target AGN and comparison stars
are not of equal brightness, and hence, the variances of comparison star DLCs are scaled by ${\omega}$ to
compensate for the larger photometric errors of fainter objects \citep[see][]{Howell88,Joshi11}.
The scaling factor ${\omega}_{j}$ is defined as
\begin{equation}
\label{eq:Omega}
{\omega}_j = \frac{\langle {\sigma}^{2}_{i, err} {\rm (AGN - ref)} \rangle}{\langle {\sigma}^{2}_{i, err} {\rm (s_j - ref)} \rangle}
\end{equation}
which is the ratio of average rms error in `target $-$ reference star' DLC to that of `comparison star $-$ reference star'
\citep[see][]{Joshi11,Ojha21}.
\
Due to the limited FoV (5$^{\prime}$.2 $\times$ 5$^{\prime}$.2) of CCD and sparse source density around our
target sources, we could use the same two comparison stars for the $F_{\rm enh}$-test that were used for
the $F^{\eta}$-test. We point out that despite using the same set of two comparison stars in the $F_{\rm enh}$-test
it enhances the power of INOV test as it does not follow the standard $F-$distribution \citep[see][]{Goyal13}.
Using equation~\ref{eq:EnFtest} we estimated $F_{\rm enh}$ for each DLC and compared it with the critical
values ($F_{\rm c}$) at 99 and 95 per cent of significance levels.
A target AGN is considered as variable (V), probable variable (PV), and non-variable (NV)
if $F_{\rm enh}$ $>$ $F_{\rm c}(0.99)$, $F_{\rm c}(0.95)$ $<$ $F_{\rm enh}$ $\leq$ $F_{\rm c}(0.99)$, and
$F_{\rm enh}$ $\leq$ $F_{\rm c}(0.95)$, respectively. In Table~\ref{tab:INOVTests}, we list  $F_{\rm enh}$ values
and corresponding INOV status in all 22 sessions.
\subsection{Amplitude of variability and duty cycle}
For each target source showing INOV we
determine peak-to-peak amplitude of INOV ($\psi$) for each DLC which is defined as
\begin{equation}
\label{eq:amp}
{\psi} = \sqrt{(A_{\rm max} - A_{\rm min})^{2} - 2{\sigma}^{2}_{\rm err}}
\end{equation}
\citep[see][]{Heidt96} where $A_{\rm max}$ and $A_{\rm min}$ are the maximum and minimum values in `AGN $-$ comparison star' DLC and ${\sigma}^{2}_{\rm err}$ = ${\eta}^{2} \langle {\sigma}^{2}_{AGN - s} \rangle$ is the mean average of rms errors of individual data points. The average rms errors are scaled up by $\eta$ = 1.54 accounting for the underestimation of errors rendered by IRAF and DAOPHOT (see Section~\ref{F-testStd}).
For a given session we compute the mean of $\psi$ obtained from the two DLCs (`AGN - S1' and `AGN - S2') related to two comparison stars.
We also computed average INOV amplitude ($\overline{\psi}$) of the sample by taking the mean of
average INOV amplitudes ${\psi}$ of only variable (`V') RL-NLS1s.
Further, we compared $\overline{\psi}$ of our sample with various sub-samples of RL-NLS1s
reported in the literature (see Table~\ref{tab:AvgINOV}).
\par
To examine the frequency of occurrence of INOV in our sample we compute duty-cycle (DC) of INOV which is defined
as
\begin{equation}
\label{eq:DC}
DC = 100 \frac{{\sum}_{i=1}^{n}R_i (1/{\Delta}t_i)}{{\sum}_{i=1}^{n}(1/{\Delta}t_i)}
\end{equation}
\citep[see][]{Romero99,Stalin04} where ${\Delta}t_i$ = ${\Delta}t_{i, observed}~(1+z)^{-1}$ is the target AGN’s redshift ($z$) corrected
time duration of the $i^{\rm th}$ monitoring session. $R_{i}$ is considered to be 1 if INOV is detected,
otherwise 0 for the $i^{\rm th}$ session. For sources monitored in more than one session we have considered only one session of longer duration.
The INOV DCs of our sample and other sub-samples are listed in Table~\ref{tab:AvgINOV}.

\section{INOV and their characteristics in our sample}
\label{sec:INOV}
\subsection{INOV detection}
Based on the $F^{\eta}$-test, only two of our sample sources (J094857.31+002225.51 and J153102.48+435637.69) exhibit INOV
wherein at least one of the two target DLCs in a session is found to be variable (see Table~\ref{tab:INOVTests}).
We note that $F_{\rm enh}$-test is more powerful than $F^{\eta}$-test, and therefore, we give preference to the results
obtained from $F_{\rm enh}$-test.
Based on $F_{\rm enh}$-test we found three sources (J094857.31+002225.51, J150506.47+032630.82
and J153102.48+435637.69) as variable and one source (J160048.75+165724.41) as probable variable.
Thus, 04/16 (25 per cent) of our sample sources exhibit INOV. Notably, all three variable sources show
INOV during one of their two monitoring sessions.
We note that all these sources are also identified as variable using ${\chi}^2$-test and ARIMA models
(see Section~\ref{sec:ARIMA}). Thus, results from different statistical
tests {\ie}$F_{\rm enh}$-test, ${\chi}^2$-test and ARIMA models are consistent with each other.
Further, we point out that the chances of detecting INOV are higher,
if a source is monitored across more than one sessions. We caution that the fraction of sources showing INOV is only a lower
limit as majority (all but six) of our sample sources are monitored for only one session.
Also, lower errors in target as well as comparison stars magnitudes can enable us to detect weak INOV.
For instance, J093323.02-001051.62 and J094857.31+002225.51 DLCs monitored on 25.01.2025 are impacted by relatively larger errors, and these can miss weak INOV, if present.
\par
We also perform multiple comparisons correction to assess false discovery rate (FDR)
({\ie} some sources might have $p$ values $<$0.01 purely by chance, even if null hypotheses (no variability)
is true) by using Benjamini-Hochberg procedure \citep{Benjamini95} that allows us to estimate $p$ values
which remain significant after controlling for FDR. The Benjamini-Hochberg procedure
assuming FDR of 0.2 finds three convincingly variable sources (J094857.31+002225.51 monitored on 09.02.2021, J150506.47+032630.82 monitored on 27.04.2025, and J153102.48+435637.69 monitored on 10.03.2021).
The Benjamini-Hochberg procedure performed with FDR = 0.2 for the ${\chi}^2$-test results also give similar
results. Therefore, we find that the most of our sources identified as variables based on $F_{\rm enh}-$test as well as
${\chi}^{2}-$test are genuine and not false positives. We also caution that Benjamini-Hochberg procedure aims to
identify FDR but stringent criterion (a low value assumed for FDR) may cause the missing of genuine sources.
%
%
\begin{table*}[ht]
\centering
\begin{minipage}{180mm}
\caption{INOV statistical results for our sample sources}
\label{tab:INOVTests}
\hskip-3.0cm
\scalebox{0.75}{
\begin{tabular}{lccccccccccc}
\hline \hline
Source & Obs. Date  &$F$-test values &INOV status & $F^{\eta}$-test & Var. & $F_{\rm enh}$ &  INOV  & {$\sqrt { \langle \sigma^2_{i,err} \rangle}$}& ${\psi}$ ($\overline{\psi}$) &  \multicolumn{2}{c}{${\chi}^2$-test}  \\ \cline{11-12}
name   &  (dd.mm.yyyy)  & {$F_1^{\eta}$},{$F_2^{\eta}$}  & $F_{\eta}$-test & $F^{\eta}_{s1-s2}$ & status & values & status  & (AGN-s)   &  (mag) & Reduced-${\chi}^2$  &  $p$-value \\
                        &            &               & (99 per cent)&         & (s1-s2) &     & (99 per cent) &       &      &     &       \\
 (1)                    &(2)         &  (3)          &    (4) & (5)   & (6)  &  (7)   &  (8)    & (9)  &  (10)   &  (11)  & (12)    \\ \hline
J080000.05$+$152326.09 & 16.11.2020 &  0.82,  0.60  & NV, NV &  0.68    &  NV   &  1.21    & NV    & 0.03  &  $...$  &  0.69, 0.55  & 0.93, 0.99    \\
J083454.90$+$553421.13 & 09.12.2020 &  0.23,  0.15  & NV, NV &  0.29    &  NV   &  1.29    & NV     & 0.02  & $...$ &  0.49, 1.15 & 0.49, 0.15 \\
J083454.90$+$553421.13 & 10.02.2021 &  0.44, 1.16   & NV, NV &  0.42    &  NV   & 1.04     & NV    & 0.01  &     $...$     & 1.11, 2.09 & 0.18, 0.11  \\
J084957.98$+$510829.08$^{\gamma}$ & 11.01.2021 &  1.15,  1.39  & NV, NV &  1.13    &  NV   &  1.02    & NV    & 0.04  &  $...$  & 1.10, 1.66 &  0.25, 0.03 \\
J090409.65$+$581527.40 & 17.12.2019 &  0.23, 0.27   &  NV, NV & 0.16    & NV  & 1.44   & NV   & 0.01  &  $...$  &
0.22, 2.39 & 1.0, 0.11 \\
J092533.67$+$021342.59 & 10.03.2021 & 0.77, 0.16  & NV, NV &  0.44   & NV  & 0.37  & NV    & 0.02  &   $...$ & 0.87, 0.16 & 0.65, 1.0 \\
J093323.02$-$001051.62 & 12.01.2021 &  0.49,  0.55  & NV, NV &  0.59    & NV   &  0.83     & NV    & 0.04  &    $...$  &  0.52, 0.55 & 0.99, 0.99 \\
J093323.02$-$001051.62 & 25.01.2025 & 0.14, 0.30   & NV, NV & 0.18     & NV   &  0.80     & NV    & 0.19 &   $...$  & 0.14, 0.27  & 1.0, 1.0  \\
J094857.31$+$002225.51$^{\gamma}$ & 09.02.2021 & 2.87, 2.74 & V, V &  0.02 &  NV   &  2.54 &  V   & 0.03  & 0.18, 0.18 (0.18) & 2.20, 2.31 & $<$0.01, $<$0.01    \\
J094857.31$+$002225.51$^{\gamma}$ & 25.01.2025 & 0.28, 0.28  & NV, NV & 0.24 & NV  &  1.19  & NV  & 0.11  &  $...$  &  0.26, 0.27  & 1.0, 1.0  \\
J103346.39$+$233220.00 & 11.03.2021 &  0.42,  0.51  & NV, NV &  1.01    &  NV   &  0.41    & NV    & 0.03  &   $...$   &  1.17, 1.80 &  0.16, 0.09 \\
J111439.00$+$324134.00 & 27.02.2020 & 1.34, 0.53    &  NV, NV & 1.27    &  NV   &  0.42    & NV    & 0.01 &  $...$ &  1.21, 0.51 & 0.15, 0.99 \\
J111752.42$+$213619.30 & 10.02.2021 & 0.19, 0.73  & NV, NV & 0.86 & NV  &  0.22   & NV    & 0.02  &   $...$ & 0.18, 0.64 & 1.0, 0.99  \\
J122039.35$+$171820.82 & 09.04.2021 &  0.48,  0.66  & NV, NV &  0.70    &  NV   &  0.69    & NV    & 0.04  &    $...$ & 0.56, 0.62  & 0.99, 0.98   \\
J130522.74$+$511640.26$\gamma$   & 10.04.2021 &  0.81,  0.70  & NV, NV &  0.60   &  NV   &  1.36    & NV    & 0.02  &   $...$  &  0.78, 0.66 & 0.92, 0.99  \\
J130522.74$+$511640.26$^\gamma$  & 21.03.2025 &  0.18, 0.10  & NV, NV  &  0.20   &  NV   &  1.02   & NV  & 0.02  &   $...$  & 0.17, 0.10   &  1.0, 1.0 \\
J143244.91$+$301435.33 & 11.04.2021 &  0.69,  0.64  & NV, NV &  0.53    &  NV   &  1.30    & NV    & 0.04  &   $...$  & 0.70, 0.64 & 0.93, 0.97 \\
J150506.47$+$032630.82$^{\gamma}$ & 09.03.2021 &  0.64,  0.87  & NV, NV &  0.37    &  NV   &  1.74  & NV  & 0.04  &  $...$  & 0.61, 1.14  &  0.97, 0.27  \\
J150506.47$+$032630.82$^{\gamma}$ & 27.04.2025 & 0.67, 0.91 & NV, NV & 0.16 & NV &  4.10 & V & 0.01 & 0.18, 0.18 (0.18)  &  3.60, 2.82   & $<$0.01, $<$0.01 \\
J153102.48$+$435637.69 & 10.03.2021 &  1.55,  3.73  & PV, V  &  0.75    &  NV   &  2.05    & V     & 0.01  & 0.09, 0.10 (0.10) &  1.53, 3.54 & $<$0.01, $<$0.01 \\
J153102.48$+$435637.69 & 30.04.2025 & 0.11, 0.10 & NV, NV & 0.06 & NV & 1.90 & PV & 0.04 & 0.26, 0.19 (0.22) & 0.17, 0.14   & 1.0, 1.0    \\
J160048.75$+$165724.41 & 11.03.2021 & 0.93, 0.89  & NV, NV &  0.50    &  NV   &  1.85    & PV    & 0.03  & 0.14, 0.12 (0.13) & 0.79, 1.17   & 0.82, 0.21      \\
 \hline
\end{tabular}
}
\\
{\bf Note.} $^{\gamma}$ : $\gamma$-ray detected NLS1s. V = variable, PV = probable variable, NV = non-variable.
The INOV amplitude (${\psi}$) for each DLC
as well as the average of two DLCs of a target source is listed only for each variable source.
\end{minipage}
\end{table*}
%
%
%
\par
Four of our RL-NLS1s (J084957.98+510829.08 aka SBS 0846+513, J094857.31+002225.51 aka PMN 0948+0022,
J130522.74+511640.26 and J150506.47+032630.82 aka PKS 1502+036)
are of particular interest due their detection in $\gamma$-ray.
We find that only two $\gamma$-ray NLS1s (J094857.31+002225.51 and J150506.47+032630.82) exhibits INOV
and remaining two $\gamma$-ray RL-NLS1s (J084957.98+510829.08 and J130522.74+511640.26) show no INOV in our
monitoring observations.
Notably, these four $\gamma$-ray RL-NLS1s were also part of the sample probed for INOV
by \cite{Ojha22} who used a minimum of two monitoring sessions each of $>$ 3.0 hours and found INOV only in J094857.31+002225.51 among these four sources.
Thus, our results on the INOV of $\gamma$-ray RL-NLS1s are broadly consistent with the findings of previous studies \citep{Ojha22}.
The INOV detection in J094857.31+002225.51 is not surprising as this source is known to be highly variable owing to its
relativistic jet viewed nearly pole-on, similar to blazars \citep[see][]{Liu10,Jiang12,Maune13,Paliya13}.
The remaining two $\gamma$-ray RL-NLS1s have been reported to show INOV in some of the previous studies
\citep[see][]{Paliya13,Paliya16}.
For instance, J084957.98+510829.08 (SBS 0846+513) is reported to have undergone a massive four magnitude optical flare
in 1975 \citep{Arp79} and nearly three magnitude optical flare in April 2013 \citep{Maune14}.
The latter optical flare was found to be associated with the increase in $\gamma$-ray, infrared, and radio fluxes.
We note that the $\gamma$-ray NLS1s are known to go through low and high state of activity with
higher occurrence of INOV in high state \citep{Paliya16}.
Therefore, absence of INOV in two $\gamma$-ray RL-NLS1s may be indicative of
their low-activity phase during our observations.
Indeed, two of our $\gamma$-ray RL-NLS1s (J084957.98+510829.08 and J130522.74+511640.26 with
their average $r$ band magnitudes 18.28 and 17.10, respectively) are found to be relatively fainter
when compared to their average magnitudes found in previous studies \citep{Maune14} as well as those
derived from the Catalina Real$-$Time Transient Survey \citep[CRTS;][]{Drake09}
long-term monitoring. For comparison, we converted median $V$ band CRTS magnitude to $r$ band using equation given in
\cite{Jester05}. Although, we note that the lack of INOV, particularly, in J130522.74+511640.26 can be also
attributed to mis-aligned and underdeveloped relativistic jets \citep{Liao15}.
\begin{figure*}[ht]
\centering
\includegraphics[angle=0, width=18cm, trim={0.0cm 0.0cm 0.0cm 0.0cm}, clip]{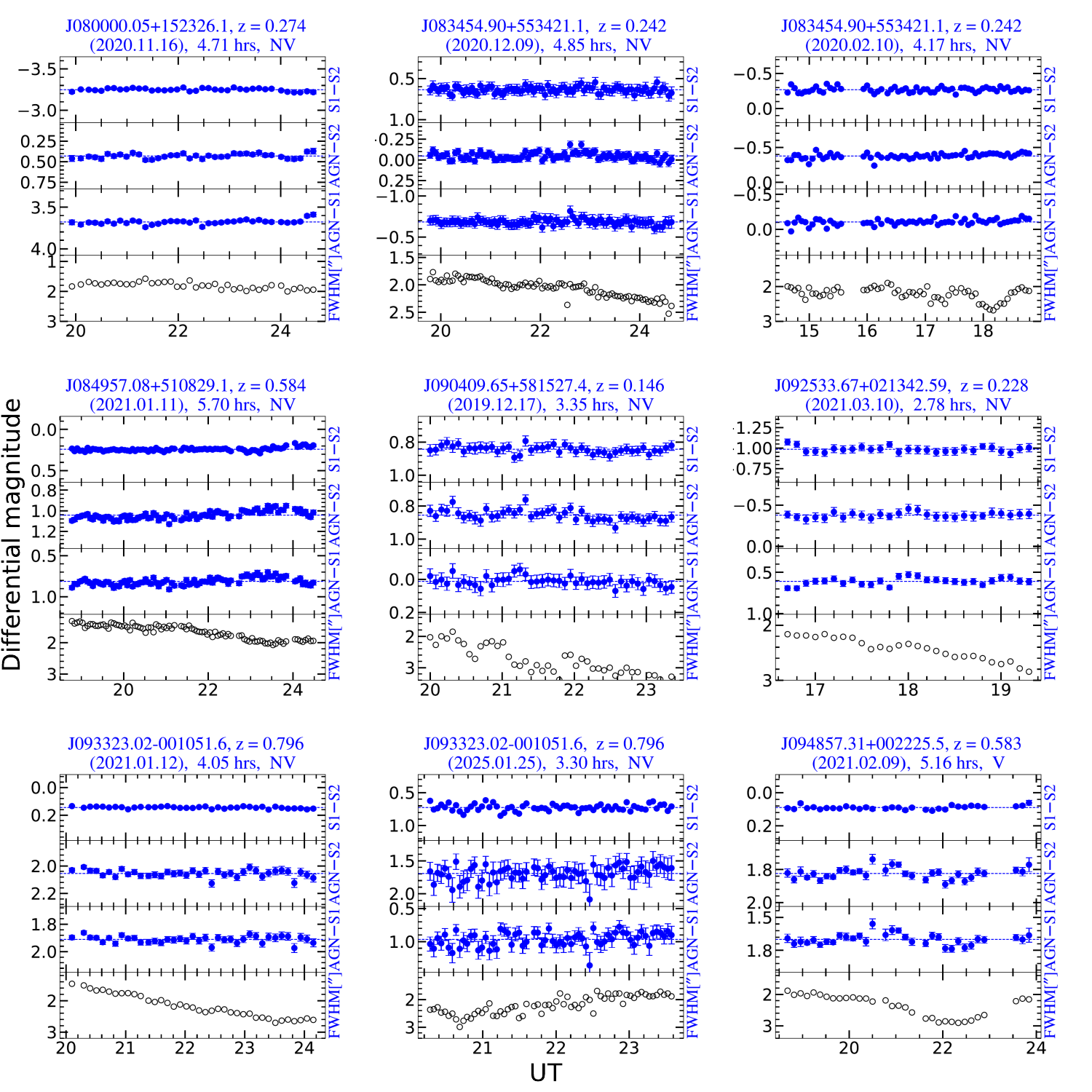}
\caption{$R$ band differential light curves (DLCs) of our $16$ RL-NLS1s. Source name, redshift, date and duration of monitoring session, and variability status are mentioned at the top of each panel. In each panel, the uppermost panel shows DLC
of two comparison stars denoted as `S1' and `S2', while next two panels show two DLCs of RL-NLS1 (AGN) relative to the two comparison stars. The lowest panel shows FWHM variations.}
\label{fig:DLCs}
\end{figure*}

\addtocounter{figure}{-1}

\begin{figure*}
\centering
\includegraphics[angle=0, width=18cm, trim={0.0cm 0.0cm 0.0cm 0.0cm}, clip]{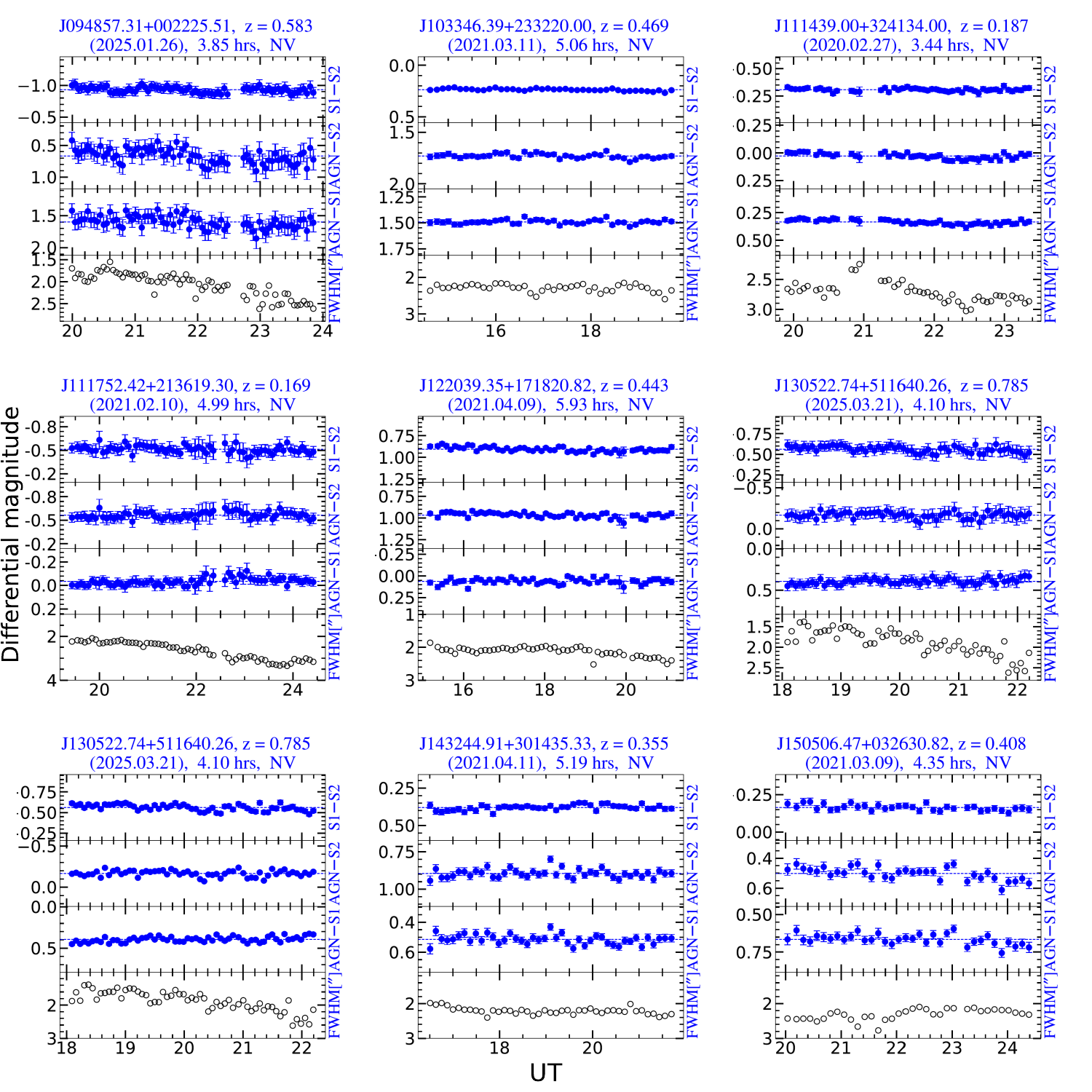}
\caption{\it Continue.}
\label{fig:DLCs}
\end{figure*}

\addtocounter{figure}{-1}

\begin{figure*}
\centering
\includegraphics[angle=0, width=13cm, trim={0.0cm 0.0cm 0.0cm 0.0cm}, clip]{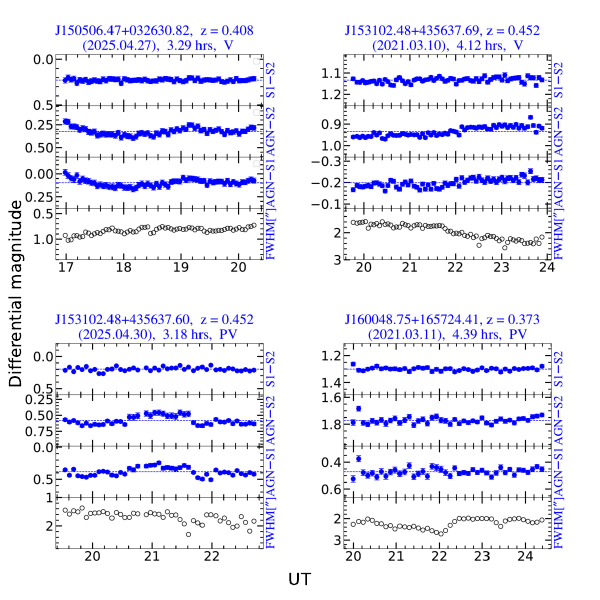}
\caption{\it Continue.}
\label{fig:DLCs}
\end{figure*}

\subsection{INOV duty cycle and amplitude}
To characterise INOV we examined duty cycle (DC; see eq~\ref{eq:amp}) and average
peak-to-peak amplitude of INOV ($\overline{\psi}$; see eq~\ref{eq:DC}). In Table~\ref{tab:AvgINOV}, we list INOV
DC and $\overline{\psi}$ using both $F^{\eta}$-test as well as $F_{\rm enh}$-test.
We prefer the results provided by the $F_{\rm enh}$-test considering its better efficacy
over the $F^{\eta}$-test in detecting INOV.
In our sample, we find DC $=$ 9 per cent and $\overline{\psi}$ $=$ 0.14 mag based on
the $F^{\eta}$-test which detected INOV in only two sources.
By using the $F_{\rm enh}$-test we detect INOV in four sources including one probable variable source and obtain DC $=$ 25 per cent
and $\overline{\psi}$ $=$ 0.16 mag. The exclusion of $\gamma$-ray NLS1s results somewhat lower INOV duty cycle and but similar average amplitude. However, considering sample biases and small size we refrain from generalizing this result.
\par
Further, we compared DC and $\overline{\psi}$ of our RL-NLS1s with other NLS1s samples,
such as $\gamma$-ray RL-NLS1s \citep{Ojha21} and jetted RL-NLS1s showing core-jet radio structures
in the VLBA observations \citep{Ojha22}, for which INOV studies were conducted using same methods and statistical
tests. Thus, we avoid bias introduced by the methods and statistical tests while comparing the average INOV properties among these samples.
\cite{Ojha21} conducted $r$ band photometric monitoring of 15 $\gamma$-ray RL-NLSy1s across 36 monitoring
sessions each lasting for a minimum of three hours and reported INOV DC around 30$-$ 40 per cent and average
amplitude of variability $\overline{\psi}$ $=$ 0.14 mag in their sample.
\cite{Ojha22} performed INOV study of a sample of 23 radio$-$bright RL-NLS1s ($S_{\rm 1.4~GHz}$ $>$ 10 mJy) with
radio-loudness parameter ($R_{\rm 1.4~GHz}$) $>$ 100.
Based on the VLBA observations they sub-divided their sample into 15 RL-NLS1s with confirmed detection of core-jet structures (jetted)
and the remaining 8 RL-NLS1s with no clear
detection of radio jets (non-jetted). They reported INOV DC of only 12$-$18 per cent in the jetted RL-NLS1s sample and nearly
no INOV (DC $=$ 0$-$5) detected in non-jetted RL-NLS1s.
The comparative analysis of INOV DC and $\overline{\psi}$ reveals that our RL-NLS1s show INOV properties
similar to $\gamma$ RL-NLS1s (see Table~\ref{tab:AvgINOV}). Notably, INOV DC of our RL-NLS1s is little lower than that
of $\gamma$-ray NLS1s sub-sample of \cite{Ojha21} which can be attributed to more monitoring sessions
per target source and inclusion of sources with smaller ($\psi$ $\geq$ 3.0 per cent) INOV amplitudes in their study.
Further, we note the INOV DC found in our sample is comparable to that known for blazars which consistently
show INOV DC $\geq$ 20 per cent \citep[e.g.,][]{Goyal13}.
Thus, high INOV DC and $\overline{\psi}$ observed in our RL-NLS1s can
be explained by the presence of relativistically beamed radio jets in most of them, if not in all.
Also, irrespective to their detection in $\gamma$-ray, our sample exhibits strikingly high DC and amplitude of variability
which in turn points to the fact that RL-NLS1s with no $\gamma$-ray detection too possess relativistic jets.
Indeed, radio properties of our RL-NLS1s also favor the presence of relativistic
jets in majority of them (see Section~\ref{sec:radio}).

\begin{table*}
\centering
\begin{minipage}{180mm}
\caption{Comparison of INOV duty cycle (DC) and average amplitude ($\overline{\psi}$) observed in different
RL-NLS1s samples}
\label{tab:AvgINOV}
\hskip-2.0cm
\begin{tabular}{lcccccccc}
\hline \hline
Sample  &  No. of   &  No. of & \multicolumn{2}{c}{$F^{\eta}$-test} &  \multicolumn{2}{c}{$F_{\rm enh}$-test} & Median log($M_{\rm SMBH}$) &  Reference \\
       & sources  & sessions & DC (\%)& $\overline{\psi}$ &   DC (\%)  &  $\overline{\psi}$  &    &   \\
 (1)    &  (2)   &   (3)    &    (4)        &  (5)    &  (6)     &   (7)    &  (8)    & (9)  \\ \hline
RL-NLS1s                          &  16 &  22  &  9   & 0.14  & 13$-$25 & 0.16  & 7.29   & This work  \\
non $\gamma$-ray RL-NLS1s         &  12 &  15  &  7   & 0.10  & 7$-$13 & 0.10$-$0.15 & 7.42    & This work  \\
${\gamma}$-ray RL-NLS1s           &  15 &  36  & 30   & 0.14  &  41      & 0.14      & $...$           & O21   \\
jetted RL-NLS1s                   &  15 &  30  & 12   & 0.11  &  18      & 0.09      & 7.72          & O22   \\
non-jetted RL-NLSy1s              &   8 &  16  & 00   & $...$   &  05      & 0.09      & 7.42          & O22   \\
\hline
\end{tabular}
\\
{\bf Note}. A range for DC and $\overline{\psi}$ is given by excluding and including probable variable (PV) sources, in case of differing values. The average peak-to-peak amplitude ($\overline{\psi}$) is estimated by taking the mean of ${\psi}$ values of all the variable sources. The jetted NLS1s are classified based on the detection of core-jet radio structures in them using VLBA observations. References $-$ O21 : \cite{Ojha21}, O22 : \cite{Ojha22}.
 \end{minipage}
 \end{table*}

\section{Radio characteristics of our RL-NLS1s}
\label{sec:radio}
To gain further insights into the nature of our RL-NLS1s
we examine their radio characteristics using publicly available radio surveys
{\eg} 144 MHz LoTSS, 150 MHz TGSS, 327 MHz Westerbork Northern Sky Survey \citep[WENSS;][]{Rengelink97}, 1.4 GHz FIRST and NVSS, and 3.0 GHz Very Large Array Sky Survey \citep[VLASS;][]{Lacy20}.
These non-simultaneous multi-frequency radio observations across 144 MHz to 3.0 GHz allow us to investigate their
radio structures, radio variability and radio SEDs.
In the following subsections we discuss their radio characteristics in detail.

\subsection{Radio structures}
To check the radio structures in our RL-NLS1s we utilize 3.0 GHz VLASS images which offer the highest
angular resolution of 2$^{\prime\prime}$.5 among all the publicly available radio surveys.
The VLASS has surveyed sky at 3.0 GHz in three different epochs during 2017 $-$ 2024 \citep{Lacy20}.
All but one (J083454.90+553421.13) of our RL-NLS1s have VLASS images in three epochs available from the
Canadian Initiative for Radio Astronomy Data Analysis (CIRADA\footnote{https://cirada.ca/}) website.
Currently, only Quick Look (QL) images are available for most of our sources. The QL images are quickly processed
without applying self-calibration and can have relatively larger noise-rms. The final data products provide single epoch (SE) images, generated by applying self-calibration, having lower noise-rms
and better dynamic range near bright sources.
Therefore, we prefer to use SE images, whenever available, else the latest QL epoch~3 images.
\par
The VLASS images show that the radio emission in all of our RL-NLS1s appear nearly unresolved.
Using {\tt JMFIT} task in the Astronomical Image Processing System (AIPS\footnote{http://www.aips.nrao.edu/index.shtml})
we find that the radio emission in all the sample sources can be simply fitted with a single elliptical Gaussian.
The {\tt JMFIT} task provides peak flux density, integrated flux density, deconvolved size and position angle of the
fitted source emission (see Table~\ref{tab:VLASSProp}).
The size of an unresolved radio source can be estimated by taking the geometrical mean of its deconvolved major axis and minor axis.
Moreover, considering significantly large uncertainties associated with the deconvolved sizes of our sources
we adopt a conservative approach by opting deconvolved size equal to the upper limit of radio size. In case deconvolved minor axis is zero
we take major axis equal to the upper limit of source size.
\par
The angular size upper limits of our sources derived from their deconvolved sizes are found to be in the range
of 0$^{\prime\prime}$.33 to 1$^{\prime\prime}$.14.
The corresponding upper limits on their physical sizes are of only a few kpc
(see Table~\ref{tab:VLASSProp}).
The radio sizes found in our RL-NLS1s seem to be consistent with previous studies suggesting mostly small-size
radio jets in NLS1s \citep{Ulvestad95,Singh18,Berton20}.
We find that the VLASS angular resolution (2$^{\prime\prime}$.5) is insufficient
to clearly resolve radio structures in our RL-NLS1s.
The VLBA/VLBI radio observations offering milli-arcsec angular resolution can be useful in resolving
the compact radio structures.
It is worth noting that some of our sources, in particular, $\gamma$-ray RL-NLS1s have been observed with VLBA and/or VLBI. For instance, VLBI observations of PMN J0948+0022 showed a compact core of high brightness
temperatures ($>$ 10$^{12}$~K) inferring the presence relativistic jets with Doppler factor 2.7$-$5.5 and
viewing angle ($\theta$) $<$ 22$^{\circ}$ \citep[see][]{Doi06}.
The VLBA observations of PMN J0948+0022 revealed a compact radio core
and a one-sided 100 pc radio jet \citep{Doi19}.
The $\gamma$-ray RL-NLS1s J084957.98+510829.0 (SBS 0846+513) and J150506.48+032630.8 (PKS 1502+036) are
also known to host relativistic jets exhibiting superluminal motion with apparent jet speed
6.6$\pm$0.6$c$ and 0.1$\pm$0.2$c$, respectively, and viewing angle $<$~20$^{\circ}$ \citep[see][]{Lister18}.
We note that the VLBI/VLBA observations confirming relativistic jets in NLS1s have been
limited mostly to $\gamma$-ray RL-NLS1s, and hence, our sample offers good candidates for VLBI observations
to detect relativistic jets in RL-NLS1s.
\subsection{Radio variability}
To investigate radio variability in our sources we used three epochs of VLASS flux densities.
We determine variability index defined as \citep[see][]{Aller92,Aller03}
\begin{equation}
\label{eq:VI}
VI = \frac{(S_{\rm max} {\minus} {\sigma}_{\rm max}) {\minus} (S_{\rm min} {\plus} {\sigma}_{\rm min})}{(S_{\rm max} - {\sigma}_{\rm max}) {\plus} (S_{\rm min} {\plus} {\sigma}_{\rm min})}
\end{equation}
where $S_{\rm max}$ and $S_{\rm min}$ are maximum and minimum flux densities of a radio source amongst all three VLASS epochs, and ${\sigma}_{\rm max}$
and ${\sigma}_{\rm min}$ are their corresponding uncertainties. The variability index for a non-variable source
is negative as the change in the flux densities at two epochs is smaller than the sum of their errors.
Thus, we list variability index only when it is a positive value.
We note that the VLASS QL epoch 1 images suffer from a systematic underestimation of flux density measurements.
With updated data processing pipeline the systematic underestimation was reduced from
10 per cent in epoch 1.1 to 3.0 per cent in epoch 1.2 (see VLASS catalog user guide\footnote{https://cirada.ca/vlasscatalogueql0}).
All our sources have epoch~1.2 QL images, and hence, we increased source flux densities by 3.0 per cent.
For determining variability index we considered only peak flux densities to avoid the effect of extended emission, if exists.
Albeit, majority of our sources show unresolved compact emission, and our results remain nearly same even
if integrated flux densities are used.
\par
Further, we also examine radio variability at 1.4 GHz by comparing the FIRST and NVSS flux densities that are measured
at two different epochs separated by a few months to several years. Due to larger beam-size (45$^{\prime\prime}$) NVSS flux densities
can include extended low-surface-brightness emission, if exists, that can be missed by the FIRST due to its smaller
beam-size (5$^{\prime\prime}$). Thus, higher flux density observed in the NVSS may not necessarily be due to variability.
However, lower flux density observed in the NVSS in comparison to the FIRST can only be attributed to variability.
In Table~\ref{tab:VLASSProp}, we list 1.4~GHz variability index for sources that show lower flux density in the NVSS
in comparison the FIRST.
\par
Using VLASS data we find that 13/16 our RL-NLS1s exhibit radio variability at 3.0 GHz with variability
index spanning across 0.01 to 0.26. There are 05/13 sources showing strong 3.0~GHz variability with $VI$ $>$0.1.
The segregation of variable sources into weakly and strongly variables may not be very meaningful
as we are using sparsely sampled light curve using only three epochs of data.
Thus, seemingly weakly variable sources may turn out to be
strongly variable, if high cadence light curves are available.
Based on the comparison of FIRST and NVSS flux densities we find that
five of our RL-NLS1s are variable at 1.4~GHz.
Not surprisingly, all these five sources also show variability at 3.0~GHz, and their
variability indices at 1.4 GHz and 3.0 GHz are broadly consistent with each other.
However, an apparently weakly variable source (J093323.02-001051.62) at 3.0 GHz turns out to be
strongly variable 1.4 GHz.
In our sample, there are three non-variable radio sources that are relatively
faint ($S_{\rm 3.0~GHz}$ $<$ 6.0 mJy).
One of our sample source J083454.90+553421.13 aka 4C55.16 lacks VLASS images at two of three epochs.
\par
Overall, our analysis shows that the majority of our RL-NLS1s are variable at radio wavelengths and
several of these show high variability index ($>$ 0.1) similar to $\gamma$-ray-loud NLS1s,
flat-spectrum radio quasars (FSRQs) and blazars.
We point out that recently, \cite{Shao25} performed multi-frequency (six bands across 2.1 GHz to 33 GHz)
long-term radio monitoring of NLS1s, and reported that, in their sample, $\gamma$-ray-loud NLS1s characterized
with compact emission, inverted or flat spectrum, exhibit stronger radio variability, while $\gamma$-ray-quiet
NLS1s often showing steep radio spectra exhibit only moderate variability.
Also, long-term monitoring of AGN at GHz frequencies has shown that, unlike other subclasses of AGN,
FSRQs and blazars are highly variable with their average 2.3~GHz variability indices found to
be 0.15 and 0.17, respectively \citep[see][]{Sotnikova22}.
Thus, high variability indices and compact radio emission in our sources again points their similarity with
$\gamma$-ray NLS1s and blazars.

\begin{table*}
\caption{Radio properties of our RL-NLS1s from all three epochs of VLASS}
\label{tab:VLASSProp}
\hskip-3.0cm
\scalebox{0.75}{
\begin{tabular}{lccccccccccc}
\hline \hline
Source &  \multicolumn{2}{c}{VLASS Epoch~1} &  \multicolumn{2}{c}{VLASS Epoch~2} &  \multicolumn{2}{c}{VLASS Epoch~3}  & \multicolumn{2}{c}{Var. Index} & ${\theta}_{\rm deconvolved}$   & size \\
name   &  $S_{\rm peak}$ &  $S_{\rm int}$  &  $S_{\rm peak}$ &  $S_{\rm int}$  &  $S_{\rm peak}$ &  $S_{\rm int}$  & $VI^{\rm 3.0~GHz}$ &  $VI^{\rm 1.4~GHz}$ &  ${\theta}_{\rm min}$ $\times$ ${\theta}_{\rm max}$, PA  &       \\   \cline{2-3}  \cline{4-5}  \cline{6-7}
       &  (mJy)          &        (mJy)    &     (mJy)       &  (mJy)          &   (mJy)         &  (mJy) &    &    &  ($^{\prime\prime}$ ${\times}$$^{\prime\prime}$, $^{\circ}$)  &  ($^{\prime\prime}$, kpc)      \\ \hline
J080000.05$+$152326.09 & 4.80$\pm$0.28 & 4.88$\pm$0.17  & 5.06$\pm$0.21$^{\rm S}$ & 5.66$\pm$0.39 & 4.72$\pm$0.22 & 5.88$\pm$0.44 & $...$    & $...$    & 0.99$\times$0.67, 9.7 &    0.81 (3.43) \\
J083454.90$+$553421.13 & $...$  &  $...$  & 7018.10$\pm$1.56 & 7238.00$\pm$2.75 & $...$ &  $...$ & $...$ & $...$ & 0.49$\times$0.37, 22.6  & 0.43 (1.63)  \\
J084957.98$+$510829.08$^{\gamma}$ & 165.78$\pm$0.18 & 174.79$\pm$0.33 & 182.03$\pm$0.13 & 188.54$\pm$0.22 & 255.62$\pm$0.20 &          267.52$\pm$0.36 &  0.21 & 0.12  & 0.57$\times$0.46, 63.9 & 0.51 (3.42)    \\
J090409.65$+$581527.45 & 5.36$\pm$0.10  & 5.97$\pm$0.20  & 4.99$\pm$0.14 & 5.22$\pm$0.24 & 4.54$\pm$0.11 & 5.10$\pm$0.21 & $...$     &  $...$  & 1.14$\times$0.0, 147.3  & $<$ 1.14 (2.92)    \\
J092533.67$+$021342.59 & 3.39$\pm$0.13 & 3.64$\pm$0.24 & 3.38$\pm$0.14 & 3.74$\pm$0.27$^{\rm S}$ & 3.29$\pm$0.13 & 3.39$\pm$0.22 &  0.06 & $...$  & 0.96$\times$0.74, 161.3 & 0.84 (3.09) \\
J093323.02$-$001051.62 &  74.46$\pm$0.16 & 77.27$\pm$0.28 & 69.29$\pm$0.19 & 70.86$\pm$0.34$^{\rm S}$ & 74.80$\pm$0.14 & 77.50$\pm$0.25 & 0.04 & 0.21  &  0.45$\times$0.34, 36.2 & 0.39 (2.98) \\
J094857.31$+$002225.51$^{\gamma}$ &  211.33$\pm$0.19 & 219.81$\pm$0.34 & 96.99$\pm$0.16 & 99.97$\pm$0.27$^{\rm S}$ & 181.24$\pm$0.16 & 186.69$\pm$0.29 &  0.37 & 0.22  & 0.66$\times$0.32, 142.7 & 0.46 (3.07) \\
J103346.39$+$233220.00 & 6.90$\pm$0.14 & 7.18$\pm$0.24 & 5.32$\pm$0.11 & 5.95$\pm$0.21 &  3.96$\pm$0.13 & 4.42$\pm$0.24 & 0.25 & $...$  & 1.24$\times$0.43, 139.0 & 0.73 (4.35) \\
J111439.00$+$324134.00 & 40.39$\pm$0.16 & 42.03$\pm$0.28 & 40.08$\pm$0.12 & 42.33$\pm$0.22 & 39.31$\pm$0.13 & 41.66$\pm$0.23 & 0.01  &   $...$  & 0.76$\times$ 0.49, 86.3 & 0.61 (1.92) \\
J111752.42$+$213619.30 & 13.42$\pm$0.13 & 13.78$\pm$0.23 & 12.57$\pm$0.11 & 12.72$\pm$0.19 & 11.74$\pm$0.12 & 12.04$\pm$0.22 & 0.06  & $...$ & 0.40$\times$0.28, 10.7 & 0.33 (0.97) \\
J122039.35$+$171820.82 & 43.99$\pm$0.14 & 46.68$\pm$0.26 & 42.69$\pm$0.13 & 46.90$\pm$0.24 & 42.71$\pm$0.13 & 46.59$\pm$0.24 & 0.01
& $...$  & 1.00$\times$0.58, 72.6 & 0.76 (4.39) \\
J130522.74$+$511640.26$^{\gamma}$ & 46.03$\pm$0.14 & 55.18$\pm$0.27 & 49.40$\pm$0.15 & 56.71$\pm$0.28$^{\rm s}$ & 47.11$\pm$0.11 &
52.25$\pm$0.22 &  0.03 & $...$ & 1.41$\times$0.45, 163.9 & 0.79 (6.03) \\
J143244.91$+$301435.33 &  21.86$\pm$0.12 & 23.15$\pm$0.22 & 20.22$\pm$0.14 & 22.20$\pm$0.26 & 22.87$\pm$0.12 & 23.65$\pm$0.21 & 0.06 &  $...$  & 0.66$\times$0.33, 109.2 & 0.47 (2.34) \\
J150506.47$+$032630.82$^{\gamma}$ & 501.36$\pm$0.25 & 522.67$\pm$0.45 & 357.14$\pm$0.42 & 397.77$\pm$0.78 & 679.68$\pm$0.28 & $...$
711.45$\pm$0.49 & 0.31 & $...$  & 0.73$\times$ 0.37, 24.3  & 0.52 (2.85) \\
J153102.48$+$435637.69 & 32.65$\pm$0.13 & 34.98$\pm$0.23 & 33.29$\pm$0.15 & 34.98$\pm$0.27 & 30.35$\pm$0.17 & 33.77$\pm$0.32 & 0.04 &    0.08 & 1.26$\times$0.57, 62.0 &  0.85 (4.94) \\
J160048.75$+$165724.41 & 21.08$\pm$0.13 & 21.91$\pm$0.22 & 36.35$\pm$0.15 & 39.48$\pm$0.28 & 25.93$\pm$0.14 & 26.57$\pm$0.25 & 0.26 &     0.26  & 0.43$\times$0.27, 177.2 & 0.34 (1.77) \\
\hline
\end{tabular}
}
{\bf Note.}  $VI^{\rm 3.0~GHz}$ : variability index at 3.0 GHz derived from VLASS flux densities. $VI^{\rm 1.4~GHz}$ :
variability index at 1.4 GHz derived using FIRST and NVSS flux densities.
The deconvolved sizes are derived from {\tt JMFIT} task using single epoch (SE) images, whenever available,
or VLASS QL~3 images. $^{\rm s}$ : values derived using SE images.
\end{table*}

\subsection{Radio SEDs}
Redio spectral energy distributions (SEDs) of RL-NLS1s can allow us to ascertain the nature of radio emission in them.
In general, radio emission in AGN can have contributions from various components {\eg} AGN driven jet-lobes, accretion disk coronal emission and on-going star-formation in their host galaxies.
The radio emission arising from star-formation processes can be composed of free-free emission (FFE) from HII regions
and synchrotron emission from supernova remnants, resulting a power law SED with index (${\alpha}$) -0.8$\pm$0.4 \citep{Gim19,An21}. The AGN radio SEDs can be characterised with steep ($\alpha$ $\leq$ -0.7) power law
if synchrotron emission is arising mainly from an optically-thin plasma associated with extended lobes and/or AGN outflows.
In misaligned AGN, radio SEDs also show a spectral turn-over at low-frequencies ($<$ 1.0 GHz)
due to synchrotron self-absorption (SSA). The spectral turn-over peak frequency (${\nu}_{\rm p}$) is supposedly
related to the radio source age such that the young radio sources show ${\nu}_{\rm p}$ $>$ 1.0 GHz
that progressively shifts towards lower frequency as radio source ages and grows larger \citep{Fanti95,ODea98,ODea21}.
In case of AGN with jets aligned close to the observer's line-of-sight, radio SED is characterised with a flat or
inverted ($\alpha$ $\geq$ -0.5) power law spectrum due to dominant optically-thick synchrotron emission from the compact-core \citep{Healey07,Massaro13a}.
\par

We examine the radio SEDs of our RL-NLS1s using their integrated flux densities at 144 MHz, 150 MHz, 327 MHz, 1.4 GHz
and 3.0 GHz. All but four of our sources have radio flux density measurements at three or more frequencies.
Four sources have flux density measurements only at 1.4 GHz and 3.0 GHz (see Table~\ref{tab:RadioProp}).
It is important to note that simultaneous observations at different frequencies are warranted to avoid the
effect of variability. Although, given the unavailability of simultaneous observations we attempt to characterise
radio SEDs of our sources using non-simultaneous data. In Figure~\ref{fig:SEDs} we show the radio SEDs of our RL-NLS1s.
We find that 14/16 of our sample sources suffer from strong variability that give rise
large breaks in their SEDs and make it impossible to constraint their overall shapes.
Two sources {\ie} J122039+171820 and J130522+511640 showing somewhat weak variability may be modelled with a simple power law with spectral
index of -0.66 and -0.51, respectively.
The in-band VLASS spectral indices available for
three of our sources (J092533.67+021342.59, J094857.31+002225.51 and J130522.74+511640.26) are also found to be
either flat or inverted (see Table~\ref{tab:RadioProp}).
Thus, we find that our RL-NLS1s are compact, flat or inverted spectrum, variable radio sources.
The compact radio size is likely to be due to the projection effect as none of our sources shows peaked radio SED
similar to young radio sources. In summary, we find that the radio characteristics of our RL-NLS1s resembles
with blazars.

\begin{table*}
\begin{minipage}{180mm}
\caption{Multi-frequency radio flux densities and spectral indices of our RL-NLS1s}
\label{tab:RadioProp}
\hskip-2.5cm
\scalebox{0.90}{
\begin{tabular}{lcccccccc}
\hline \hline
Source                 & $S_{\rm 144~MHz}^{\rm LoTSS}$ &  $S_{\rm 150~MHz}^{\rm TGSS}$ & $S_{\rm 327~MHz}^{\rm WENSS}$ & $S_{\rm 1.4~GHz}^{\rm FIRST}$ & $S_{\rm 1.4~GHz}^{\rm NVSS}$ & $S_{\rm 3.0~GHz}^{\rm VLASS}$ & ${\alpha}_{\rm 150~MHz}^{\rm 3.0~GHz}$ & ${\alpha}_{\rm 3.0~GHz}$  \\
name                   &   (mJy)         &   (mJy)           &      (mJy)        &   (mJy)           &   (mJy)           &     (mJy)         &     &  in-band     \\
(1)             &  (2)      & (3)    &   (4)      &       (5)      &      (6)  &      (7)  &   (8)   &    \\  \hline
J080000.05$+$152326.09 &   $...$   &  $<$24.5  &  $...$   & 6.74$\pm$0.14     &  7.2$\pm$0.5      & 5.88$\pm$0.44  &  $>$-0.48 &  $...$      \\
J083454.90$+$553421.13 & 9273.8$\pm$7.6 & 11926.6$\pm$1192.7& 9184$\pm$4.7 & 8360.63$\pm$1.16 & 8283.1$\pm$248.5 & 7238.00$\pm$2.75  & -0.17$\pm$0.03  &   $...$      \\
J084957.98$+$510829.08$^{\gamma}$ & 193.7$\pm$1.2  & 267.3$\pm$27.2    & 202$\pm$4.2  & 350.53$\pm$0.15 & 266.3$\pm$8.0 & 267.52$\pm$0.36 & 0.00$\pm$0.03 &   $...$    \\
J090409.65$+$581527.45 & 24.0$\pm$0.4  &  $<$24.5  &  $<$18.0   & 6.37$\pm$0.12 &   6.2$\pm$0.4    & 5.10$\pm$0.21  &  $>$-0.52 & $...$                  \\
J092533.67$+$021342.59 &  $...$   &  $<$24.5   &  $...$  & 6.28$\pm$0.15    &  6.4$\pm$0.5     & 3.39$\pm$0.22  &  $>$-0.66 & -0.52$\pm$0.14           \\
J093323.02$-$001051.62 &  $...$  & 85.6$\pm$9.4  &  $...$  & 105.58$\pm$0.15   &   66.8$\pm$2.0   & 77.50$\pm$0.25 &  0.03$\pm$0.03
&    $...$    \\
J094857.31$+$002225.51$^{\gamma}$ &  $...$ & 68.6$\pm$12.5 &  $...$  & 111.46$\pm$0.15   &   69.5$\pm$2.1    & 186.69$\pm$0.29 &  0.33$\pm$0.06 & 0.25$\pm$0.10          \\
J103346.39$+$233220.00 &   $...$   &  27.2$\pm$5.4     &   $...$    &  9.20$\pm$0.14    &  8.7$\pm$0.5      & 4.42$\pm$0.24  & -0.61$\pm$0.07 &    $...$    \\
J111439.00$+$324134.00 & 227.8$\pm$2.9 &  154.4$\pm$16.0   &    241$\pm$3.4    &  107.77$\pm$0.12  & 109.9$\pm$3.3     & 41.66$\pm$0.23 &  -0.44$\pm$0.03  &     $...$       \\
J111752.42$+$213619.30 & $...$ & $<$24.5  & $...$  &  13.08$\pm$0.14  & 13.2$\pm$0.6      & 12.04$\pm$0.22 &  $>$-0.24  &   $...$      \\
J122039.35$+$171820.82 &   $...$ &  393.8$\pm$39.7 &  $...$ & 87.55$\pm$0.15   & 95.5$\pm$2.9      & 46.59$\pm$0.24 &  -0.71$\pm$0.03  &     $...$      \\
J130522.74$+$511640.26$^{\gamma}$ & 276.4$\pm$2.6 &  273.2$\pm$27.6   &     206$\pm$2.6   & 86.94$\pm$0.15  & 87.0$\pm$2.6      & 52.25$\pm$0.22 &  -0.55$\pm$0.03 & -0.64$\pm$0.10  \\
J143244.91$+$301435.33 & 335.5$\pm$1.8 &  226.1$\pm$23.2   &     195$\pm$4.1   & 49.98$\pm$0.14   & 50.6$\pm$1.6      & 23.65$\pm$0.21 &  -0.75$\pm$0.03 &    $...$       \\
J150506.47$+$032630.82$^{\gamma}$ &  $...$  &  120.0$\pm$15.4   &  $...$  & 380.49$\pm$0.15  &  394.8$\pm$11.9   & 711.45$\pm$0.49 &  0.60$\pm$0.04 &  $...$    \\
J153102.48$+$435637.69 & 37.9$\pm$0.3 &   $<$24.5         &     28$\pm$4.0    & 55.46$\pm$0.15    &  45.4$\pm$1.4     & 33.77$\pm$0.32 &  $>$0.11 &    $...$      \\
J160048.75$+$165724.41 &   $...$    &   $<$24.5         &     $...$      & 30.16$\pm$0.15   &  17.1$\pm$0.7     & 26.57$\pm$0.25 &  $>$0.03 &    $...$      \\
\hline
\end{tabular}
}
\\
{\bf Note.} An upper limit on flux density is used for sources lying within a survey region without detection.
The upper limit on 150 MHz flux density is based on 7$\sigma$ = 24.5 mJy detection limit of TGSS Alternative Data Release 1
(ADR1) catalog \citep{Intema17}. The upper limit on 327 MHz flux density is based on 5$\sigma$ = 18 mJy detection limit in WENSS
\citep{Rengelink97}. The VLASS flux densities are from the latest epoch~3.
150 MHz $-$ 3.0 GHz spectral indices (${\alpha}_{\rm 150~MHz}^{\rm 3.0~GHz}$) are derived using non-simultaneous observations
and they are impacted by variability.
\end{minipage}
\end{table*}

\section{Discussion}
\label{sec:Discussion}
\subsection{RL-NLS1s as scaled down version of FSRQs}

\begin{figure*}
\centering
\includegraphics[angle=0, width=8.0cm, height=5.5cm, trim={0.0cm 0.0cm 0.0cm 0.0cm}, clip]{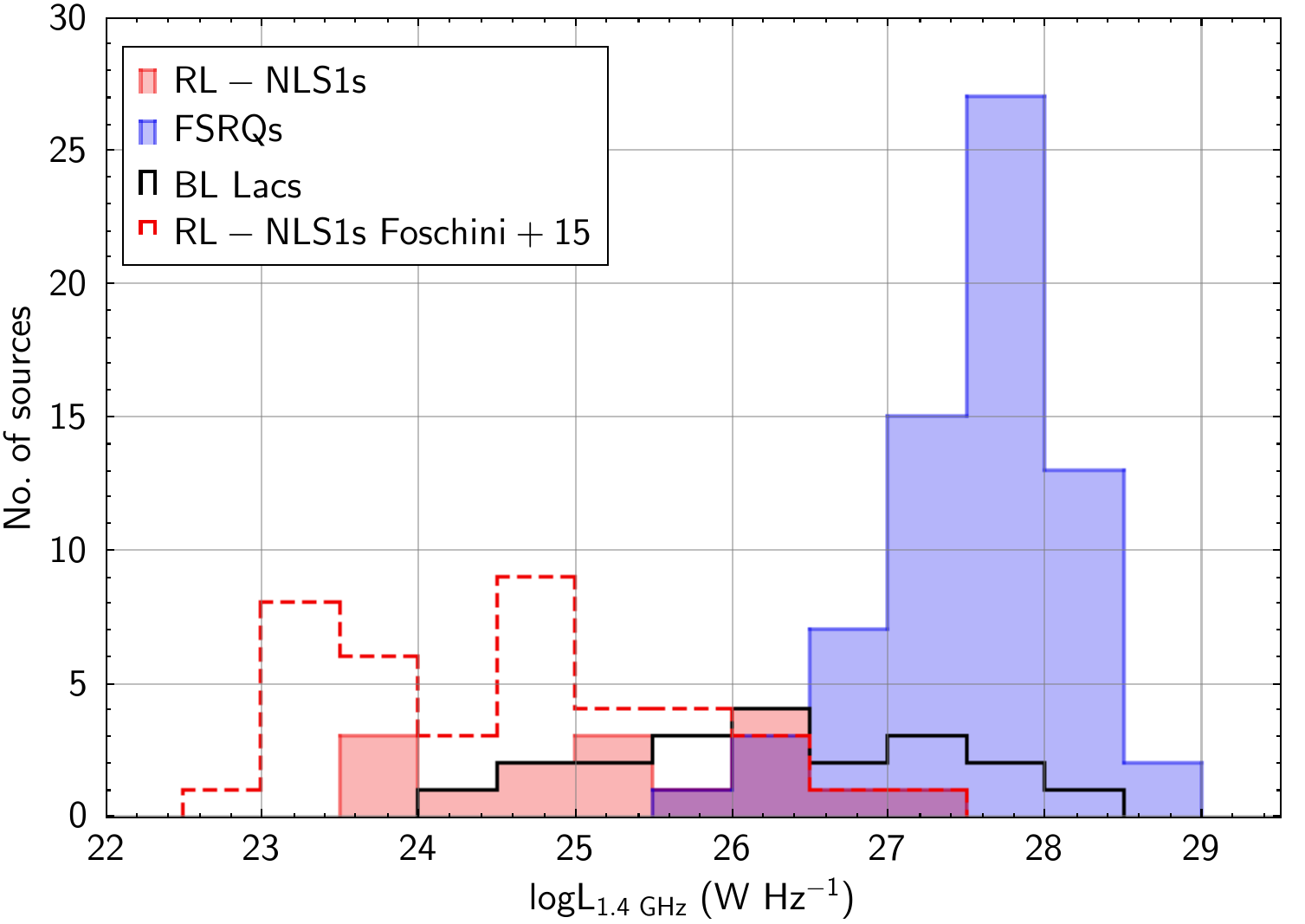}
\includegraphics[angle=0, width=8.0cm, height=5.5cm, trim={0.0cm 0.0cm 0.0cm 0.0cm}, clip]{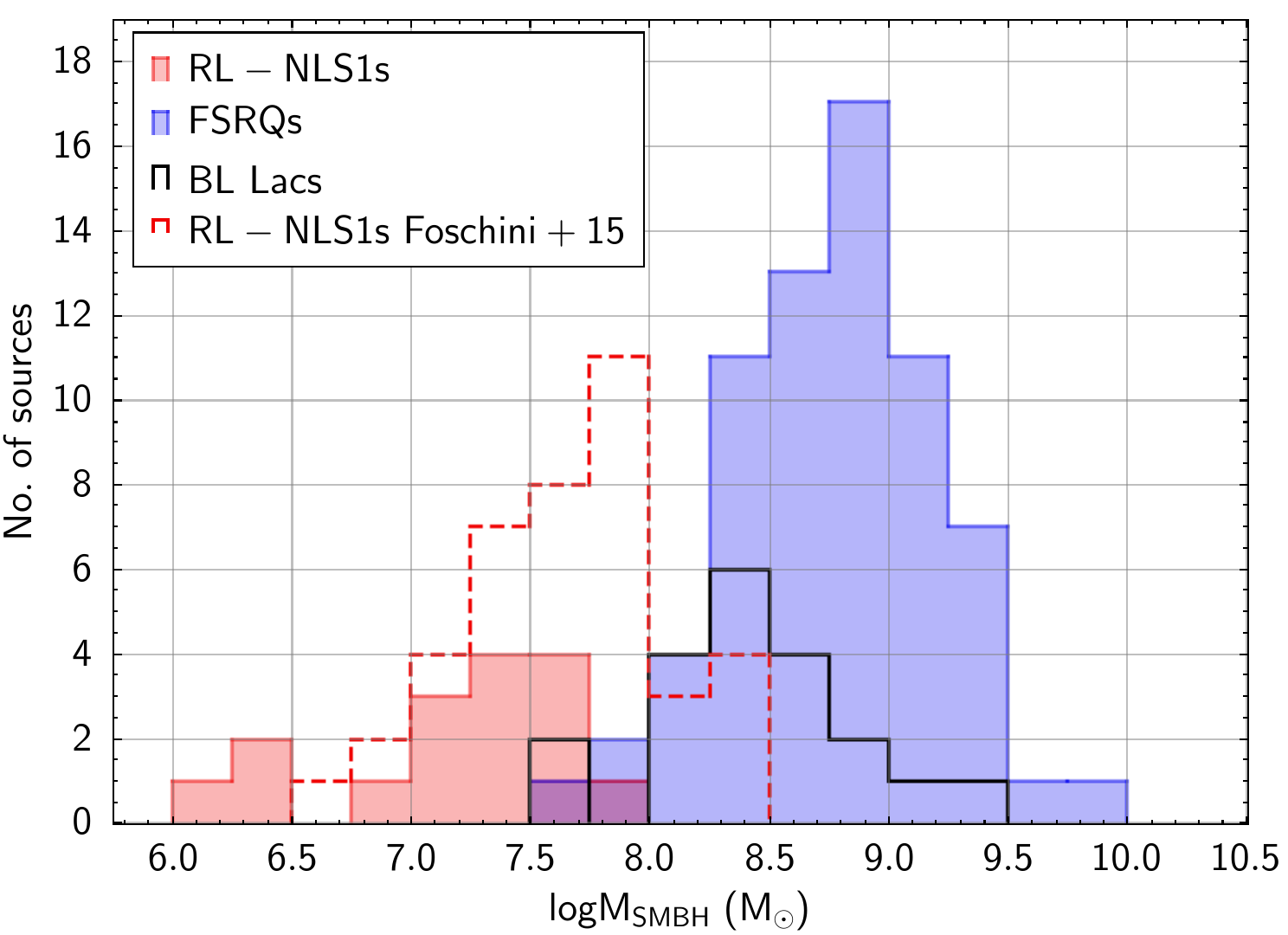}
\caption{{\it Left panel} : 1.4 radio luminosity distributions of different samples of RL-NLS1s and blazars. {\it Right panel} : Black hole mass distributions of different samples of RL-NLS1s and blazars.}
\label{fig:HistSMBHLradio}
\end{figure*}

\begin{table*}
\caption{Comparison of 1.4 GHz radio luminosities, SMBH masses and redshifts of RL-NLS1s, FSRQs and BL-Lacs}
\label{tab:SMBH}
\hskip-2.0cm
\begin{tabular}{cccccccccccc}
\hline
Type  & No. of & \multicolumn{3}{c}{log$L_{\rm 1.4~GHz}$ (W~Hz$^{-1}$)} & \multicolumn{3}{c}{log$M_{\rm SMBH}$ (M$_{\odot}$)} & \multicolumn{3}{c}{Redshift ($z$)} & Reference \\ \cline{3-5} \cline{6-8} \cline{9-11}
             & sources &  range    & median & SD  & range       & median & SD & range         & median & SD &           \\ \hline
RL-NLS1s     &  16 & 23.51$-$27.07 & 25.34 & 1.02 & 6.03$-$7.97 & 7.29 & 0.52 & 0.146$-$0.796 & 0.390 & 0.20 & 1   \\
RL-NLS1s (FS) &  40 & 22.97$-$27.12 & 24.58 & 1.04 & 6.72$-$8.49 & 7.71 & 0.42 & 0.061$-$0.918 & 0.344 & 0.20 & 2 \\
FSRQs        &  68 & 25.57$-$28.65 & 27.62 & 0.59 & 7.59$-$9.80 & 8.80 & 0.43 & 0.158$-$2.534 & 0.978 & 0.54 & 3      \\
BL-Lacs      &  20 & 24.08$-$28.06 & 26.23 & 1.10 & 7.65$-$9.26 & 8.41 & 0.39 & 0.030$-$1.254 & 0.303 & 0.38 & 3     \\
\hline
\end{tabular}
{\bf Note.} FS : flat-spectrum RL-NLS1s. SD : standard deviation of the parameter distribution.
References - 1 : This work; 2 : \cite{Foschini15}; 3 : \cite{Zhang24}.
\end{table*}

To probe the question whether our RL-NLS1s are indeed similar to blazars, in particular, FSRQs,
we compare some of their fundamental parameters such as
radio luminosity and black hole mass. In jetted AGN, radio luminosity can be regarded as an indicator of the jet power
\citep{Cavagnolo10,Godfrey13}, and black hole mass supposedly plays a key role in powering the jet \citep{Heinz03}.
To get a better statistics on the comparison between RL-NLS1s and blazars we also include a sample of flat spectrum RL-NLS1s
reported in \cite{Foschini15}, which show radio properties similar to our RL-NLS1s.
For blazars, we take the sample from \cite{Zhang24} for which SMBH mass estimates are available in the literature.
We derive 1.4 radio luminosity of blazars using the FIRST data, whenever available, or NVSS data,
and apply $K$-correction assuming a flat ($\alpha$ = 0) radio spectral index.
We find that 3.0 GHz VLASS images for several extremely bright blazars are either unavailable or severely affected by strong artefacts.
We keep the two subclasses of blazars {\ie} FSRQs and BL Lacs objects, separately, as our objective is to investigate
similarity between RL-NLS1s and FSRQs. Unlike BL-Lacs, radio jets in FSRQs make slightly larger
angle (5$^{\circ}$$-$10$^{\circ}$) {\it w.r.t.} observer's line-of-sight \citep[e.g.,][]{Urry95,Scarpa97} and they show prominent
emission lines in their optical spectra, an observational characteristic similar to RL-NLS1s.
\par
In Figure~\ref{fig:HistSMBHLradio} (left panel) we show 1.4 GHz radio luminosity distributions of RL-NLS1s and blazars.
It is clear that our RL-NLS1s are systematically less luminous in compared to both FSRQs as well as BL Lacs.
For our RL-NLS1s, 1.4 GHz radio luminosities are distributed across 23.51 W~Hz$^{-1}$ $\leq$ $L_{\rm 1.4~GHz}$ $\leq$ 27.07 W~Hz$^{-1}$
with a median value of (log$L_{\rm 1.4~GHz}^{\rm median}$) 25.34 W~Hz$^{-1}$, while FSRQs radio luminosities span across
25.57 W~Hz$^{-1}$ $\leq$ log$L_{\rm 1.4~GHz}$ $\leq$ 28.65 W~Hz$^{-1}$ with a median value of 27.62 W~Hz$^{-1}$ (see Table~\ref{tab:SMBH}).
It can be noted that despite having seemingly different radio luminosities, RL-NLS1s and FSRQs radio luminosity distributions
overlap at log$L_{\rm 1.4~GHz}$ $>$ 25 W~Hz$^{-1}$ (see Figure~\ref{fig:HistSMBHLradio}, (left panel)).
The flat-spectrum RL-NLS1s from \cite{Foschini15} reinforce the same trend.
We also point out that, in compared to FSRQs, RL-NLS1s are at systematically lower redshifts (see Table~\ref{tab:SMBH}).
Keeping the fact in mind that RL-NLS1s and FSRQs show similarity in terms of INOV, radio varaibilty, radio SEDs, compact radio emission,
we conclude that RL-NLS1s are low$-z$, low-luminosity analogs of FSRQs.
\par
\begin{figure}
\centering
\includegraphics[angle=0, width=8.5cm, trim={1.0cm 9.0cm 0.5cm 9.0cm}, clip]{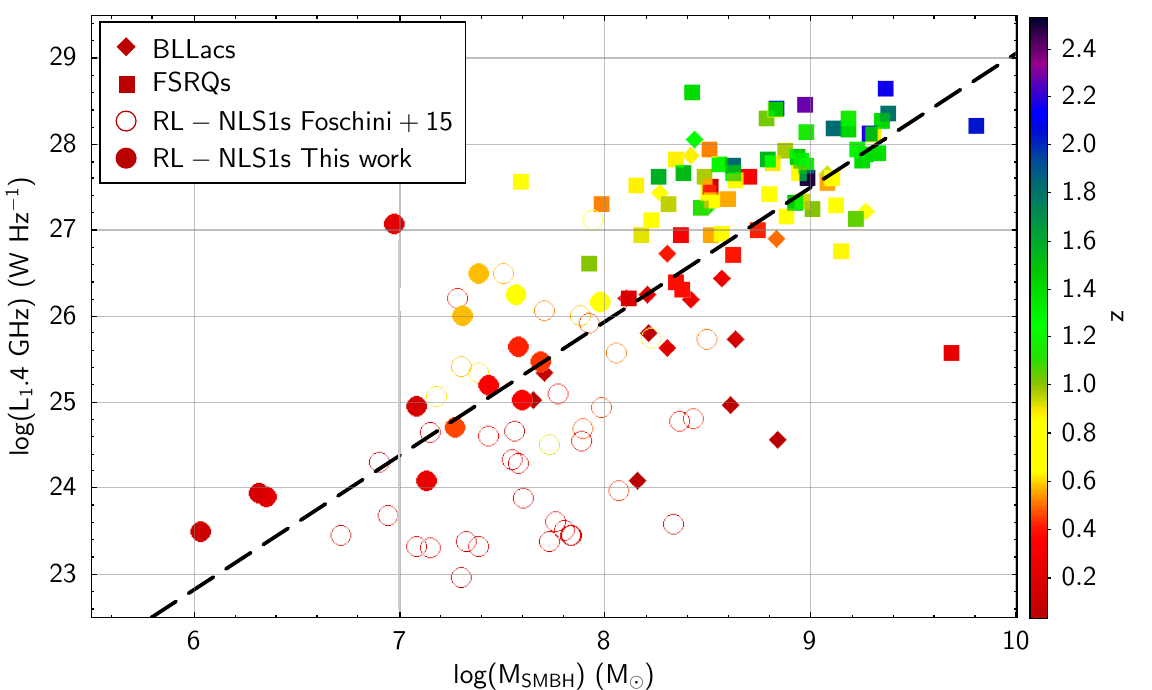}
\caption{The black hole mass versus radio luminosity plot for RL-NLS1s and blazars.}
\label{fig:SMBHVsLradio}
\end{figure}

The comparison of SMBH mass distributions reveal that NLS1s possess SMBHs of systematically lower masses in compared
to FSRQs and BL-Lacs (see Figure~\ref{fig:SMBHVsLradio}, right panel).
Our RL-NLS1s have SMBH masses in the range of 1.07 $\times$ 10$^{6}$ M$\odot$ to 9.33 $\times$ 10$^{7}$ M$\odot$ with
a median value of 1.95 $\times$ 10$^{7}$ M$\odot$. The SMBHs in FSRQs have masses in range of
3.89 $\times$ 10$^{7}$ M$\odot$ to 6.31 $\times$ 10$^{9}$ M$\odot$ with a median value of 6.30 $\times$ 10$^{8}$ M$\odot$
(see Table~\ref{tab:SMBH}). Thus, in compared to FSRQs, RL-NLS1s are powered by, on average, one-order-of-magnitude less
massive SMBHs, which is consistent with previous studies \citep{Yuan08,Dong12,Foschini20}.
The SMBH mass estimates are based on the virial method employing line-widths of broad-line region (BLR) emission lines.
The systematically less massive SMBHs found in RL-NLS1s is attributed to the smaller line-widths
(FWHM of H$\beta$ $<$ 2000~km~s$^{-1}$) seen in their optical spectra.
Although, geometrical effects may also give rise smaller line width resulting lower SMBH mass estimates \citep{Decarli08,Calderone13,Liu16}.
\par

We further explore the comparison between RL-NLS1s and FSRQs by plotting SMBH mass ($M_{\rm SMBH}$) versus
1.4 GHz radio luminosity ($L_{\rm 1.4~GHz}$) (see Figure~\ref{fig:SMBHVsLradio}).
As expected, RL-NLS1s and FSRQs occupy distinctively different regions in this plot which simply confirms that,
in comparison to FSRQs, RL-NLS1s possess less massive SMBHs and less powerful jets. We find that $L_{\rm 1.4~GHz}$ and $M_{\rm SMBH}$
are correlated with Spearman Rank correlation coefficient of 0.55, once all the jetted AGN (RL-NLS1s, FSRQs and BL Lacs) are
taken together.
The $M_{\rm SMBH}$ and $L_{\rm 1.4~GHz}$ correlation is best fitted with a linear regression equation
log$L_{\rm 1.4~GHz}$ $=$ 1.56$\times$$M_{\rm SMBH}$ + 13.46. The observed correlation between $M_{\rm SMBH}$
and $L_{\rm 1.4~GHz}$ demonstrates that jet power scales up with the black hole mass,
at least among jetted AGN which is consistent with previous studies \citep{Heinz03,Foschini17}.
We note that, unlike FSRQs, RL-NLS1s show a much larger scatter, especially when flat-spectrum RL-NLS1s from \cite{Foschini15}
are included. The large scatter shown by flat-spectrum RL-NLS1s could be attributed to the fact that they possess less powerful radio
jets as evident from their low radio luminosity and low radio-loudness {\ie} several sources with $R$ $<$ 100.
Notably, our RL-NLS1s show less scatter as they are extremely radio-loud ($R$ $>$ 100) and jet dominated AGN,
similar to FSRQs. The scatter seen in the $M_{\rm SMBH}$ and $L_{\rm 1.4~GHz}$ correlation plot
can also be taken as an indication that other fundamental parameters such as accretion rate (accretion mode) and
black hole spin may also play a role in defining the jet power \citep[see][]{Fan19,Zhang24}.
Further, we need to account for the beaming effect which requires jet speed and viewing angle measurements that remain unavailable
for the most of RL-NLS1s. In nutshell, $M_{\rm SMBH}$ and $L_{\rm 1.4~GHz}$ correlation observed for RL-NLS1s and FSRQs
together favours the conformity between two sub-classes of jetted AGN wherein jet power depends on the black hole mass.
\subsection{Origin of INOV}
The origin of INOV in beamed jetted AGN is generally attributed to mechanisms operating in their relativistic jets.
In literature, various mechanisms have been proposed to explain INOV.
The widely believed scenarios invoke shocks propagation or bulk injection of energetic particles in the jet causing
variability on short time-scales \citep[{e.g.,}][]{Marscher85,Spada01,Giannios10}.
The relativistically enhanced small fluctuations from turbulent plasma motion in the jet can also be a viable
mechanism \citep{CalafutWiita15}. The jet helicity or jet precession is also capable of producing variable
emission \citep{Camenzind92,GopalKrishna92,Abraham2000,Britzen18}. In addition to various jet driven mechanisms one
cannot rule out the possibility of INOV produced via fluctuations in the accretion disc \citep[see][]{Mangalam93}.
Although, accretion disc fluctuations may not be large and frequent enough to explain the strong and rapid INOV.
The stark differences found between the INOV properties of beamed AGN (blazars, $\gamma$-ray NLS1s) and unbeamed
AGN (low polarised radio quasars, radio quiet AGN) offer evidence in favor of jet driven INOV
\citep[see][]{Goyal12,Ojha22}. In fact, high INOV DC (22$-$28 per cent) observed even in seemingly radio-quiet AGN
with low-mass SMBHs (10$^{6}$~$M_{\odot}$) is also attributed to small-scale relativistic jets \citep[see][]{Gopal23}. Thus, noting the high duty cycle (25 per cent) and large average amplitude ($\overline{\psi}$ = 0.16)
of INOV in our RL-NLS1s, similar to blazars, we infer that relativistic jets are likely to cause INOV observed in them. Although, we caution that contribution from mechanisms such as magnetic reconnection in the magnetosphere of
SMBH causing INOV in non-jetted AGN \citep{DeGouveia10,Ripperda22}, cannot be ruled out completely.

\section{Summary and conclusions}
\label{sec:conclusions}
We investigated INOV and radio properties of a sample of 16 extremely radio-loud RL-NLS1s ($R_{\rm 1.4~GHz}$ $>$ 100)
among which four are detected in $\gamma$-ray.
Using $F_{enh}$ statistical test performed on their $R$ band photometric DLCs we detected INOV only in 04/16 RL-NLS1s.
The average INOV duty cycle and amplitude ($\overline{\psi}$) are 25 per cent and 0.16 mag, respectively,
similar to those found in $\gamma$-ray NLS1s and blazars.
Also, regardless of their detection in $\gamma$-ray, our RL-NLS1s continue to exhibit INOV of
strikingly high duty cycle and amplitude that can be attributed to the Doppler boosting of optical emission from a relativistic jet.
\par
Using 3.0 GHz VLASS and auxiliary radio data (144 MHz LoTSS, 150 MHz TGSS, 325 MHz WENSS and 1.4 GHz FIRST and NVSS)
we investigated radio sizes, radio varaibilty and radio spectral shapes of our sample sources. We note that
all our RL-NLS1s appear unresolved in the VLASS images offering highest angular resolution of 2$^{\prime\prime}$.5
amongst all the radio surveys.
The VLBA/VLBI radio observations of three of our sample sources available in the literature
show parsec-scale compact core-jet structures and
confirm the presence of relativistic jets viewed at small angles ($<$ 20$^{\circ}$).
In the remaining RL-NLS1s, compact emission inferred from the VLASS images can be explained by two
scenarios - (i) intrinsically small-size jets viewed at any angle, (ii) large-size jets viewed at small angle.
We favour the second scenario considering radio variability and lack of peaked-spectrum sources.
Based on the comparison of VLASS flux densities from three epochs we find radio variability in 13/16 our RL-NLS1s
with 3.0 GHz variability index spanning across 0.01 to 0.26.
Five of these sources show strong variability with index $>$ 0.1.
We point out that high variability indices ($>$ 0.1) seen in several of our RL-NLS1s are similar to those found in
gamma-ray NLS1s and blazars.
The radio SEDs generated from multi-frequency (144 MHz $-$ 3.0 GHz) but non-simultaneous radio data infer
flat or inverted spectral shapes for our RL-NLS1s.
The in-band VLASS spectral indices too, whenever available, exhibit flat or inverted spectral shapes.
\par
Further, by using $L_{\rm 1.4~GHz}$ $-$ $M_{\rm SMBH}$ correlation plot we demonstrate that jet power scales up with
the black hole mass for both RL-NLS1s and FSRQs in a similar fashion, despite RL-NLS1s possessing
low-luminosity and low-mass SMBHs. Noting the similarity between RL-NLS1s and FSRQs in terms of
their INOV properties, radio characteristics (sizes, variability and spectral shapes) and $L_{\rm 1.4~GHz}$ $-$ $M_{\rm SMBH}$ correlation, we conclude
that our RL-NLS1s can be regarded as low-luminosity (powered by low-mass SMBHs) analogs of FSRQs.

\begin{acknowledgments}
We thank the reviewer for insightful comments that helped to improve the manuscript.
The research work carried out at Physical Research Laboratory (PRL) is funded by the Department of Space,
Government of India. PK, AKD and VO thank PRL for their fellowships. We acknowledge the use of
data collected from MFOSC$-$P and CMOS imager on 1.2m telescope at Mt. Abu Observatory.
We thank Dr. Mudit K. Srivastava, Dr. Vipin Kumar, Prof. Shashikiran Ganesh, Mrs. Deekshya R. Sarkar and local support
staff at Mt. Abu observatory for their help during observations.
This work makes use of data provided by the National Radio Astronomy Observatory, a facility of the
National Science Foundation operated under cooperative agreement by Associated Universities, Inc.
\end{acknowledgments}

\begin{contribution}
VS designed the project, sample and wrote the manuscript. PK carried out observations
with MFOSC-P on 1.2m Mt. Abu telescope and reduced the data. AKD carried out observations with CMOS imager on 1.2m Mt. Abu telescope
and reduced the data. VO was involved in performing analysis and statistical tests.


\end{contribution}

%

\facilities{Mt Abu observatory:1.2m, VLA, GMRT, LOFAR}

\software{IRAF \citep{Tody86},
                   AIPS \citep{Greisen90}
          }


\appendix

\begin{longtable*}[c]{ccccccccc}
\caption{Details of monitoring observations of our RL-NLS1s} \\
\hline \hline
Source  &   Date       & RA (J2000)  & DEC (J2000) &  $g$    & $r$     & $g-r$  &  Time  &  No. of data   \\
name    &  (dd.mm.yy)    & (h m s)     & (d m s)     & (mag)   & (mag)   & (mag)  &  (hours)   &    points             \\
(1)     & (2)          & (3)         & (4)         & (5)     & (6)     & (7)    &   (8)  &    (9)           \\
\hline
J080000.05+152326.09 &  16.11.2020  & 08 00 00.05 &+15 23 26.1  &   19.38 &  18.57 &  0.81  &  4.71  &  38  \\
S1                    &              & 07 59 55.32 &+15 23 54.0  &   15.22 &  14.89 &  0.33  &        &      \\
S2                    &              & 08 00 01.38 &+15 20 37.8  &   18.51 &  18.23 &  0.28  &        &      \\
J083454.90+553421.13 &  09.12.2020  & 08 34 54.90 &+55 34 21.1  &   18.40 &  17.19 &  1.21  &  5.04  &  86  \\
S1                    &              & 08 35 00.83 &+55 35 23.6  &   20.53 &  18.92 &  1.61  &        &      \\
S2                    &              & 08 35 08.15 &+55 35 43.4  &   17.17 &  16.51 &  0.66  &        &       \\
J083454.90+553421.08 &  10.02.2021  & 08 34 54.90 &+55 34 21.1  &   18.40 &  17.19 &  1.21  &  3.99  &  57   \\
S1                    &              & 08 34 40.29 &+55 33 21.5  &   18.99 &  17.62 &  1.37  &        &       \\
S2                    &              & 08 35 08.15 &+55 35 43.4  &   17.17 &  16.51 &  0.66  &        &       \\
J084957.98+510829.08$^{\gamma}$ &  11.01.2021  & 08 49 57.98 &+51 08 29.1  &   18.91 &  18.28 &  0.63  &  5.70  &  87   \\
S1                    &              & 08 50 12.62 &+51 08 08.0  &   19.54 &  18.06 &  1.48  &        &       \\
S2                    &              & 08 50 16.73 &+51 09 23.3  &   18.19 &  17.80 &  0.39  &        &       \\
J090409.65+581527.45 &  17.12.2019  & 09 04 09.65 &+58 15 27.4  &   17.66 &  16.63 &  1.03  &  3.35  &  44   \\
S1                    &              & 09 03 47.06 &+58 13 55.1  &   18.09 &  17.08 &  1.01  &        &       \\
S2                    &              & 09 04 08.21 &+58 16 03.3  &   16.95 &  15.78 &  1.17  &        &       \\
J092533.67+021342.59 &  10.03.2021  & 09 25 33.67 &+02 13 42.6  &   19.76 &  18.08 &  1.68  &  2.78  &  27   \\
S1                    &              & 09 25 31.78 &+02 13 16.8  &   18.44 &  17.71 &  0.73  &        &       \\
S2                    &              & 09 25 32.88 &+02 14 03.9  &   19.34 &  18.70 &  0.64  &        &       \\
J093323.02-001051.62 &  12.01.2021  & 09 33 23.01 &-00 10 51.6  &   18.59 &  18.56 &  0.03  &  4.05  &  38   \\
S1                    &              & 09 33 31.02 &-00 11 11.9  &   17.63 &  17.12 &  0.51  &        &       \\
S2                    &              & 09 33 34.25 &-00 10 46.5  &   18.51 &  17.11 &  1.40  &        &       \\
J093323.02-001051.62 &  25.01.2025  & 09 33 23.01 &-00 10 51.6 & 18.65 & 18.62 & 0.03  &  3.35  &  67   \\
S1                  &              & 09 33 31.02 & -00 11 11.9 & 17.65 &  17.15   &  0.51   &        &       \\
S2                  &              & 09 33 34.25 & -00 10 46.5 & 18.51 &  17.10  &   1.40   &        &       \\
J094857.31+002225.51$^{\gamma}$ &  09.02.2021  & 09 48 57.31 &+00 22 25.5  & 18.59 &  18.43 &  0.16  &  5.16  &  32   \\
S1                    &              & 09 49 00.44 &+00 22 34.9 & 18.27 & 16.84 &  1.43  &        &       \\
S2                    &              & 09 48 57.29 &+00 24 18.2 & 17.00 & 16.58 &  0.42  &        &       \\
J094857.31+002225.51$^{\gamma}$ &  26.01.2025  & 09 48 57.31 & +00 22 25.5 & 18.67  & 18.50 &  0.16  &  3.65  &  73   \\
S1                  &              & 09 49 00.44 & +00 22 34.9 & 18.29  & 16.81 &  0.12  &        &       \\
S2                  &              & 09 48 57.29 & +00 24 18.2 & 17.02  & 16.60 &  0.42  &        &       \\
J103346.39+233220.00 &  11.03.2021  & 10 33 46.39 &+23 32 20.0  &   18.66 &  18.51 &  0.15  &  5.08  & 42    \\
S1                    &              & 10 33 53.65 &+23 34 05.8  &   17.76 &  17.11 &  0.65  &        &       \\
S2                    &              & 10 33 43.04 &+23 30 57.4  &   17.36 &  16.63 &  0.73  &        &       \\
J111439.00+324134.00 &  27.02.2020  & 11 14 39.00 &+32 41 34.0  &   18.94 &  17.13 &  1.81  &  3.44  & 51    \\
S1                    &              & 11 14 34.85 &+32 42 21.4  &   16.39 &  15.90 &  0.49  &        &       \\
S2                    &              & 11 14 42.65 &+32 40 23.0  &   18.50 &  17.51 &  0.91  &        &       \\
J111752.42+213619.30 &  10.02.2021  & 11 17 52.42 &+21 36 19.3  &   18.66 &  17.54 &  1.12  &  4.99  & 62    \\
S1                    &              & 11 17 44.72 &+21 35 52.9  &   18.64 &  17.43 &  1.21  &        &       \\
S2                    &              & 11 17 59.41 &+21 36 29.4  &   18.80 &  17.49 &  1.31  &        &       \\
J122039.35+171820.82 &  09.04.2021  & 12 20 39.35 &+17 18 20.8  &   18.22 &  18.16 &  0.06  &  5.93  &  55   \\
S1                    &              & 12 20 38.33 &+17 18 41.4  &   19.37 &  18.02 &  1.35  &        &       \\
S2                    &              & 12 20 37.83 &+17 19 11.7  &   17.46 &  16.84 &  0.62  &        &       \\
J130522.74+511640.26$^{\gamma}$ &  10.04.2021  & 13 05 22.74 &+51 16 40.3  &   17.29 &  17.10 &  0.19  &   5.58     &  75  \\
S1                    &              & 13 05 10.91 &+51 17 17.0  &   18.23 &  16.78 &  1.45  &    &  \\
S2                    &              & 13 05 20.14 &+51 17 49.4  &   14.33 &  13.80 &  0.53  &        &     \\
J130522.74+511640.26$^{\gamma}$ &  21.03.2025  & 13 05 22.74 &+51 16 40.3 &  17.34 & 17.16 & 0.19       &  4.10  &  59   \\
S1                  &              & 13 05 10.91 &+51 17 17.0 &  18.23 &  16.78 &  1.45 &        &       \\
S2                  &              & 13 05 20.14 &+51 17 49.4 &  14.34 &  13.81 &  0.54 &        &       \\
J143244.91+301435.33 &  11.04.2021  & 14 32 44.91 &+30 14 35.3  &   18.81 &  18.56 &  0.25  &   5.19     &  43      \\
S1                    &              & 14 32 40.54 &+30 14 08.1  &   18.05 &  17.57 &  0.48  &    &    \\
S2                    &              & 14 32 38.83 &+30 13 26.5  &   17.83 &  17.19 &  0.64  &        &       \\
J150506.47+032630.82$^{\gamma}$ &  09.03.2021  & 15 05 06.47 &+03 26 30.8  &   18.64 &  18.22 &  0.42  &   4.35     &  35      \\
S1                    &              & 15 05 09.59 &+03 27 36.3  &   19.27 &  18.10 &  1.17  &    &    \\
S2                    &              & 15 05 07.50 &+03 26 17.2  &   19.10 &  18.16 &  0.94  &        &       \\
J150506.47+032630.8  & 27.04.2025 & 15 05 06.47 & +03 26 30.8 & 18.64 & 18.22 & 0.42 & 3.29  & 65      \\
 S1                  &              & 15 05 00.17  & +03 25 48.30 & 18.05 & 17.56 & 0.49  &    &    \\
 S2                  &              & 15 05 12.79 &  +03 28 34.96 & 17.82 & 17.30 & 0.52  &   &    \\
J153102.48+435637.69 & 10.03.2021 & 15 31 02.48 & +43 56 37.7 & 17.21 & 17.09 & 0.12  & 4.12 & 67 \\
S1                    &              & 15 31 06.16 &+43 53 47.9  &   17.94 &  17.32 &  0.62  &        &       \\
S2                    &              & 15 31 06.76 &+43 55 57.9  &   16.84 &  16.17 &  0.67  &        &       \\
J153102.48+435637.7  & 30.04.2025  & 15 31 02.48 & +43 56 37.7 & 17.21 & 17.09 & 0.12 &  3.18  &  44   \\
 S1                  &             & 15 31 00.68 & +43 54 25.93 & 16.78 & 16.23 & 0.55  &     &  \\
 S2             &              & 15 31 16.80 & +43 58 02.63 &  16.30 & 15.87 & 0.43  &     &  \\
J160048.75+165724.41 &  11.03.2021  & 16 00 48.75 &+16 57 24.4  &   18.28 &  18.09 &  0.19  & 4.39   &  42   \\
S1                    &              & 16 00 51.63 &+16 57 32.9  &   18.31 &  17.71 &  0.60  &        &       \\
S2                    &              & 16 00 54.89 &+16 58 31.2  &   17.69 &  16.57 &  1.12  &        &       \\
\hline
\\
\label{tab:log}
\end{longtable*}
\section{Autoregressive integrated moving average (ARIMA) model}
\label{sec:ARIMA}
The $F$-tests treat data as a set of independent measurements rather than a time-series.
Therefore, to validate our INOV results we also use Auto-Regressive Integrated Moving Average
(ARIMA) models
that consider each DLC as a time-series. In ARIMA models, a time-series is considered as a linear combination
of three components - Auto-Regression (AR), Integration (I) and Moving Average (MA) and can be expressed as
\begin{equation}
\label{eq:ARIMA}
    F_t = \mu + \sum_{i=1}^{p} \alpha_{i}F_{t-i} + \sum_{j=1}^{q} \beta_{j}\epsilon_{t-j} + \epsilon_t
\end{equation}
where, $F_t$ is a time-series represented as the sum of mean (${\mu}$), AR term, MA term,
and error ($\epsilon_{t}$) term \citep{Feigelson18}. $\alpha$ and $\beta$ are coefficients in AR and MA terms, respectively.
The AR component accounts for the relationship between a present observation with
a certain number of past (lagged) observations. The integration component makes time-series
stationary ({\ie} mean, variance, and autocorrelation remain constant over time).
In case of non-stationary data, we make data stationary by differencing method {\ie} subtracting
consecutive data points. The MA component models the relationship between the current observation and
the residual errors of moving average of past observations.
In general, ARIMA model is written as ARIMA(p,d,q), where `p', `d' and `q' refer to the order of AR component
(the number of lag observations considered), degree of differencing
(number of times the data are differenced) and the order of MA component
(size of the moving average window), respectively.
\par
To apply ARIMA models we first check whether data are stationary using Augmented Dickey-Fuller
\citep[ADF;][]{Dickey79} statistical test and make time-series data stationary by applying iterative differencing, if required
and obtain `d'.
Thereafter, we compute `p' and `q' by analysing partial autocorrelation function (PCF) and autocorrelation
function, respectively. We obtain Akaike Information Criterion (AIC) values
for different orders (0 $\leq$ p $\leq$ 3, 0 $\leq$ q $\leq$ 3) of ARIMA(p,d,q)
models and chose the best model based on the lowest AIC value.
We consider a DLC to be variable (non-constant (NC)) only if $\Delta$AIC = AIC(optimum model) - AIC(constant model) = $\leq$ -10.
In Table~\ref{tab:ARIMAResults} we list the best fitted parameters of ARIMA models for all the DLCs.
We identify four variable sources (J094857.31+002225.51 monitored on 25.01.2025, J111439.00+324134.00, J150506.47+032630.82 monitored on 27.04.2025, J153102.48+435637.69 on both monitoring sessions) where both DLC are not fitted with a constant model and
comparison stars DLC shows no variation {\ie} fitted with constant model. We find that ARIMA models results are consistent with
those from $F$-test and ${\chi}^2$-test (see Section~\ref{sec:INOV}).

\begin{table*}[ht]
\centering
\begin{minipage}{180mm}
\caption{The best fitted parameters of ARIMA models}
\label{tab:ARIMAResults}
\hskip-2.5cm
\begin{tabular}{lcccccccccc}
\hline
Source name & Obs. Date & \multicolumn{3}{c}{S1-S2} & \multicolumn{3}{c}{AGN-S1} & \multicolumn{3}{c}{AGN-S2} \\ \cline {3-5} \cline{6-8} \cline{9-11}
 &  (dd.mm.yyyy) & (p,d,q)  & AIC  & $\Delta$AIC & (p,d,q) & AIC & $\Delta$AIC & (p,d,q) & AIC  &  $\Delta$AIC  \\
\hline
J080000.05$+$152326.09 & 16.11.2020 & (1,0,0) &  -205.55  & -6.61 &  (1,0,0)   &  -166.86   &  -5.50  & (1,0,0)  & -162.96 &  -3.61 \\
J083454.90$+$553421.13 & 09.12.2020 & (0,0,0) &  -325.09 &  0 & (1,0,0) & -340.921   & -0.593 & (3,1,3)  & -295.99  & -1.51  \\
J083454.90$+$553421.13 & 10.02.2021 & (0,0,0) &  -235.38 &  0 & (0,0,0) & -214.70  & 0 & (0,0,0) & -222.69  & 0 \\
J084957.98$+$510829.08$^{\gamma}$ & 11.01.2021 & (0,1,1) & -455.98 & $-39.24^{\rm NC}$ & (0,1,1) & -331.52 & $-40.18^{\rm NC}$ & (0,1,1)  & -333.64  & $-55.60^{\rm NC}$  \\
J090409.65$+$581527.40 & 17.12.2019 & (0,0,0) &  -189.40 & 0 & (0,1,1) & -167.01  & 4.65 & (0,1,1) & -157.41 & -2.78 \\
J092533.67$+$021342.59 & 10.03.2021 & (0,0,1) & -103.71  & -0.01 & (1,0,0) & -93.99 & -2.81 & (1,0,0) & -109.425  & -0.958 \\
J093323.02$-$001051.62 & 12.01.2021 & (0,0,0) & -249.79  & 0 & (0,0,0) & -154.85 & 0 & (0,0,0)  & -150.36  &  0    \\
J093323.02$-$001051.62 & 25.01.2025 & (0,0,0) & -201.06  & 0 & (0,0,0) & -98.80   &  0 & (0,0,0) & -96.92  & 0  \\
J094857.31$+$002225.51$^{\gamma}$ & 09.02.2021 & (0,0,0) & -180.88 & 0 & (1,0,0) & -114.47 & -3.42 & (2,0,0) & -111.57  & -1.53  \\
J094857.31$+$002225.51$^{\gamma}$ & 25.01.2025 & (0,1,1) & -248.97 & -6.52 & (0,1,1) & -140.03 & $-11.21^{\rm NC}$ & (0,1,1) & 124.47  & $-19.02^{\rm NC}$ \\
J103346.39$+$233220.00 & 11.03.2021 & (1,0,0) & -268.98 & $-12.12^{\rm NC}$ & (0,0,0) & -204.63 & 0 & (1,0,0) & -201.855  & -4.176  \\
J111439.00$+$324134.00 & 27.02.2020 & (0,0,0) & -282.36 & 0 & (0,1,1) & -278.67 & $-29.42^{\rm NC}$ & (0,1,1) & -247.76 & $-21.10^{\rm NC}$ \\
J111752.42$+$213619.30 & 10.02.2021 & (0,0,0) & -228.73 &  0 & (0,1,1) & -244.26 & -11.21 & (1,0,1) & -248.18 & $-4.57^{\rm NC}$ \\
J122039.35$+$171820.82 & 09.04.2021 & (3,0,0) & -246.46 & -9.78 & (0,0,0) & -222.99 & 0 & (1,0,1) & -232.52  &  -2.40 \\
J130522.74$+$511640.26$^\gamma$ & 10.04.2021 & (2,0,0) & -506.57 & -2.45 & (1,0,1)  & -421.20  & $-16.88^{\rm NC}$ & (2,0,0) & -454.08  &  -6.91   \\
J130522.74$+$511640.26$^\gamma$ & 21.03.2025 & (2,1,1) & -219.88  & -5.12 & (1,0,1) & -231.79 & -2.87 & (0,0,0) & -219.50  & 0    \\
J143244.91$+$301435.33 & 11.04.2021 & (1,0,0) & -222.87 & -1.1 & (0,0,0) & -172.02 & 0 & (0,0,0)  & -181.12  & 0   \\
J150506.47$+$032630.82$^{\gamma}$ & 09.03.2021 & (0,0,0) & -173.40 & 0 & (0,0,0) & -132.82 & 0  & (1,0,0)  & -121.16  &  -1.32  \\
J150506.47$+$032630.82$^{\gamma}$ & 27.04.2025 & (0,0,0) & -376.56 & 0 & (1,1,0) & -305.43 & $-64.37^{\rm NC}$ & (2,0,0)  & -298.69  & $-65.27^{\rm NC}$  \\
J153102.48$+$435637.69 & 10.03.2021 & (1,0,0) & -405.94 & -0.26 & (0,1,1) & -357.91 & $-24.16^{\rm NC}$ & (0,1,1) & -381.34  & $-60.43^{\rm NC}$ \\
J153102.48$+$435637.69 & 30.04.2025 & (0,0,0) & -181.99 & 0 & (1,1,0) & -132.37 & $-18.12^{\rm NC}$ & (1,1,0) & -146.86  & $-35.80^{\rm NC}$ \\
J160048.75$+$165724.41 & 11.03.2021 & (0,0,0) & -247.66 & 0 & (0,0,1) & -178.21 & -0.21 & (0,0,0) & -193.45 &  0  \\
 \hline
\end{tabular}
\\
{\bf Note.} The best fitted parameters (p,d,q) are listed for each DLC (`S1-S2', `AGN-S1' and `AGN-S2') of each source. Akaike Information Criterion (AIC) values are given for the best-fitted ARIMA model.
A DLC is treated variable if constant model fails to give the best fit
and $\Delta$AIC = AIC(best model) - AIC(0,0,0) $<$ -10. The variable cases are marked with `NC'.
\end{minipage}
\end{table*}

\begin{figure*}
\centering
\includegraphics[angle=0, width=5.5cm, trim={0.0cm 0.0cm 0.0cm 0.0cm}, clip]{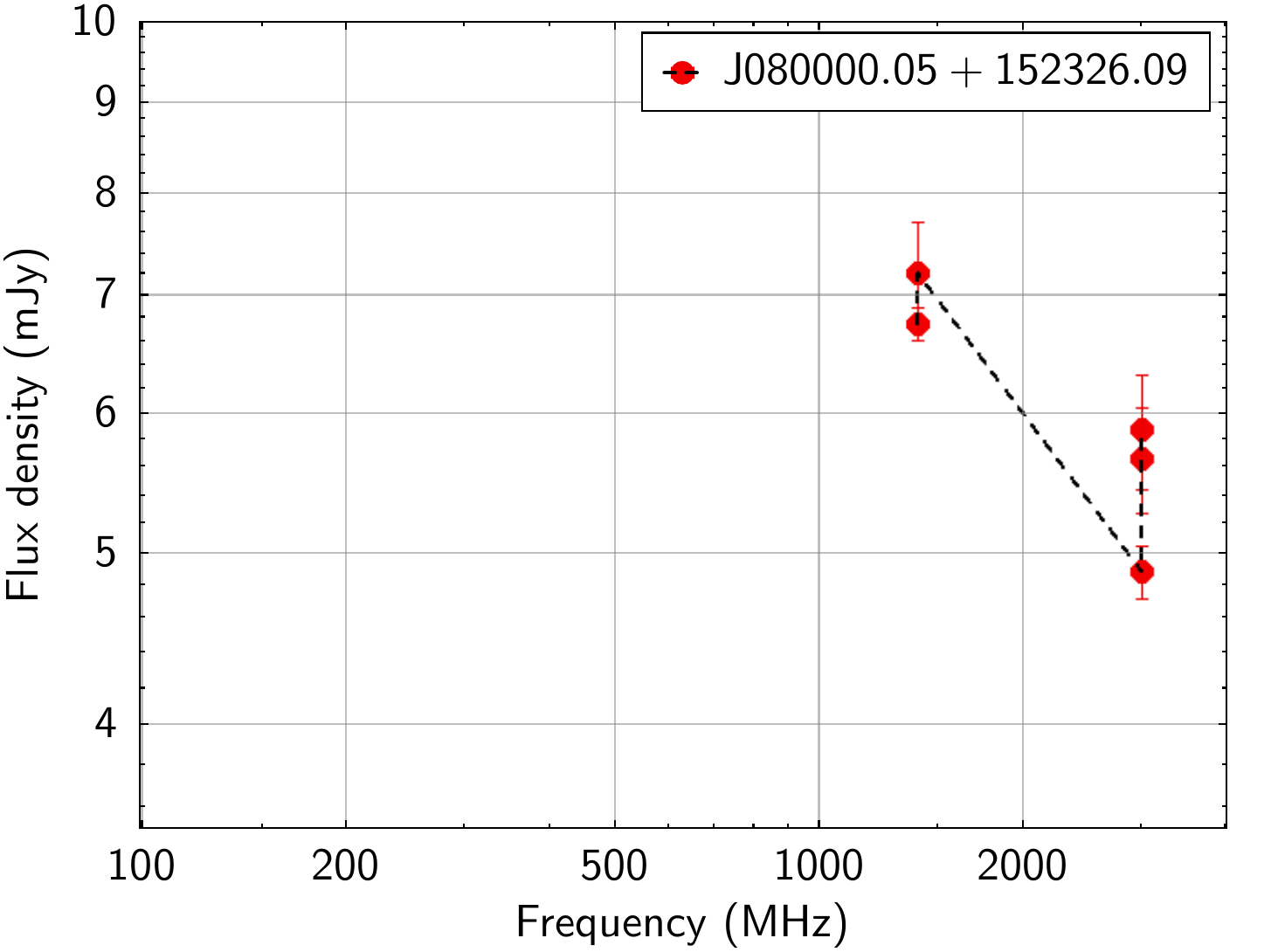}
\includegraphics[angle=0, width=5.5cm, trim={0.0cm 0.0cm 0.0cm 0.0cm}, clip]{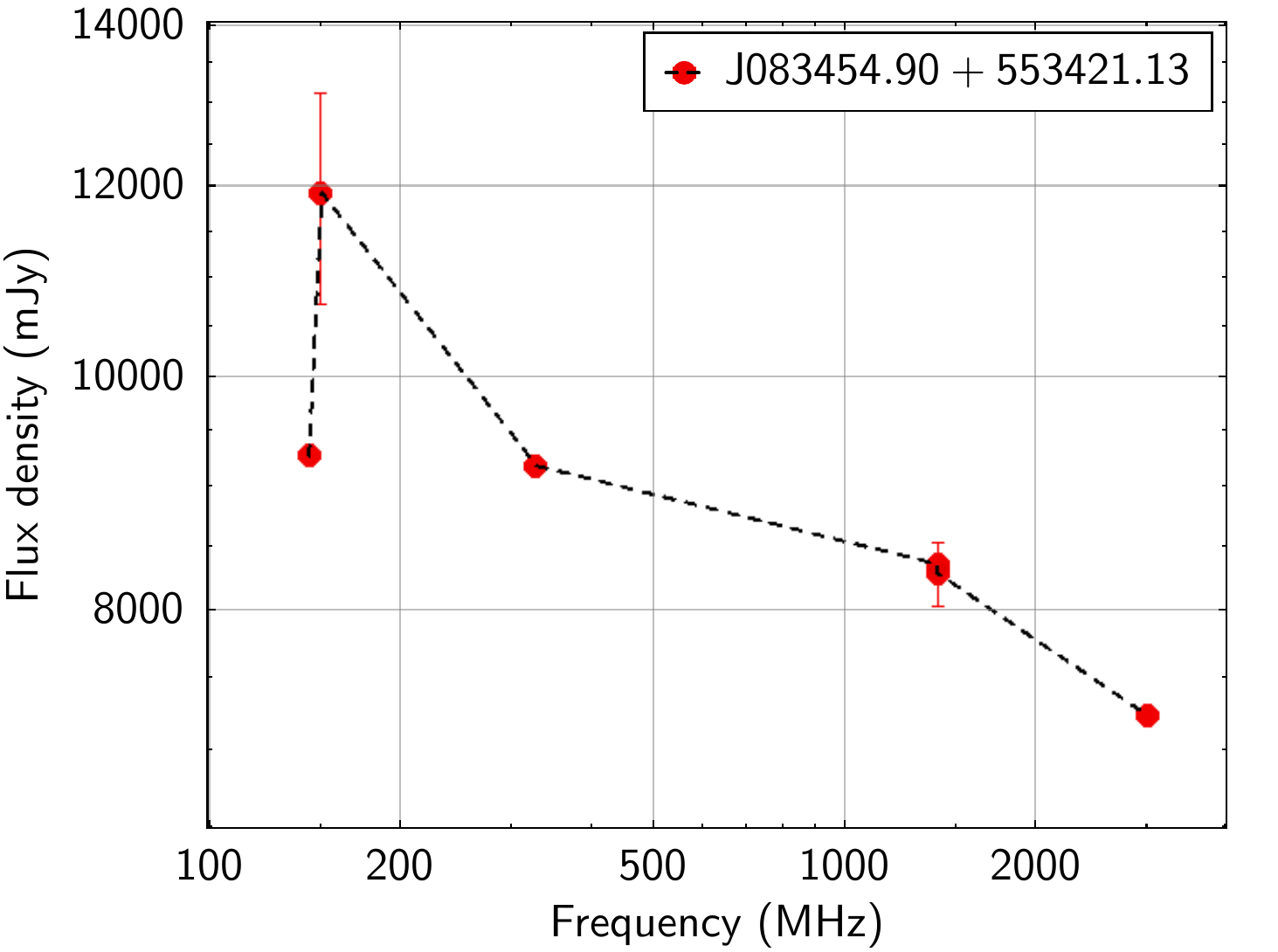}
\includegraphics[angle=0, width=5.5cm, trim={0.0cm 0.0cm 0.0cm 0.0cm}, clip]{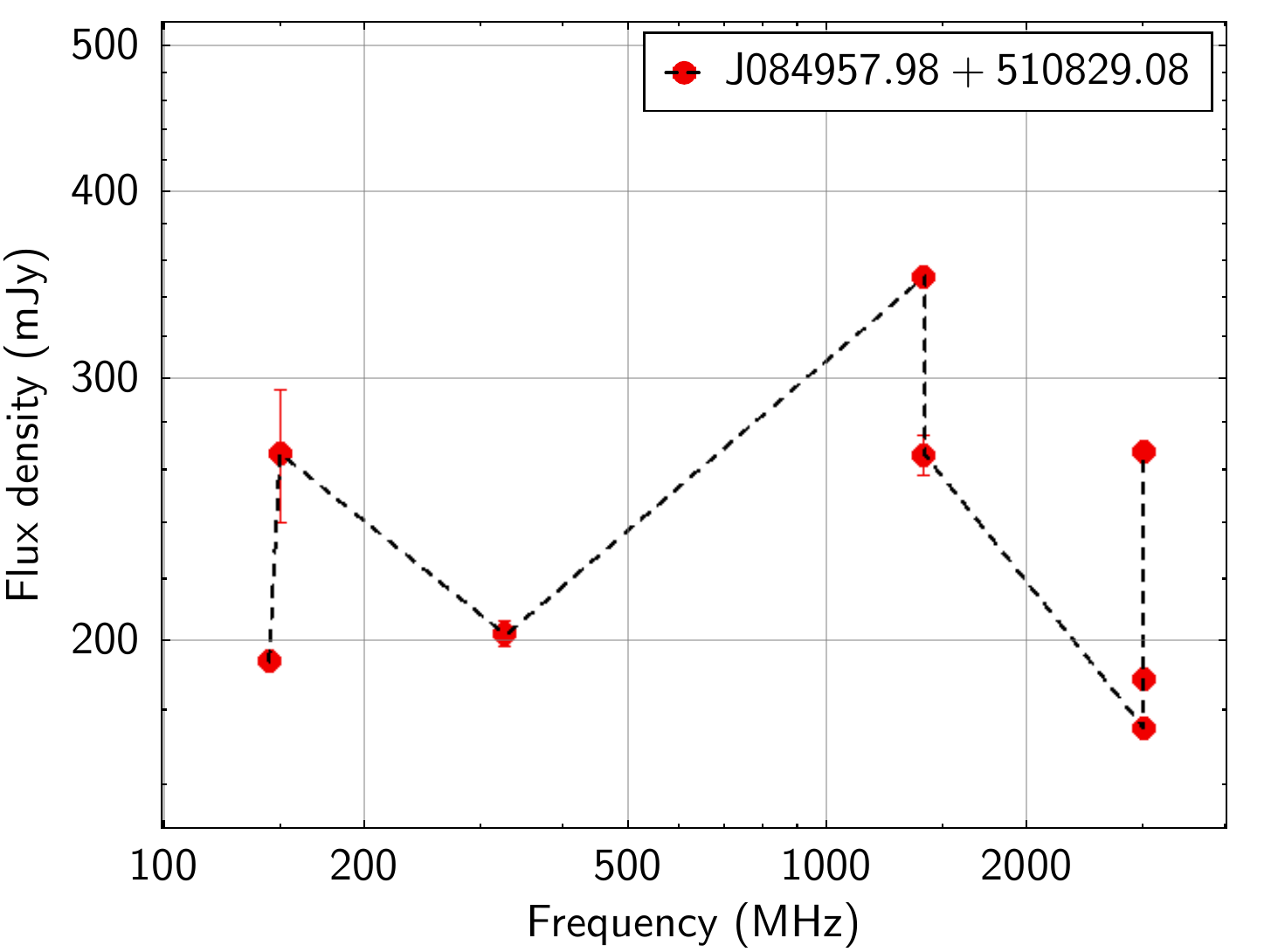}
\includegraphics[angle=0, width=5.5cm, trim={0.0cm 0.0cm 0.0cm 0.0cm}, clip]{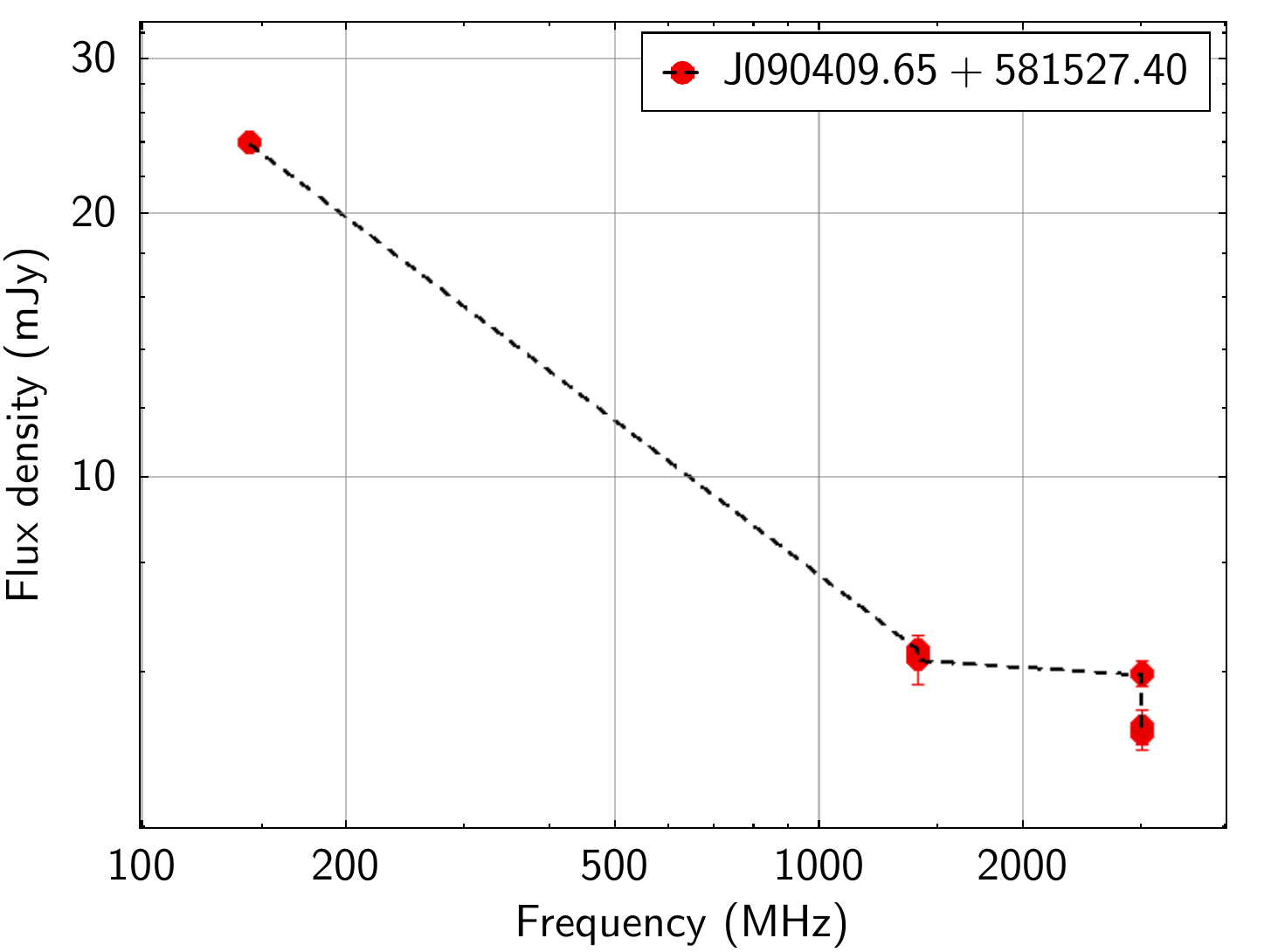}
\includegraphics[angle=0, width=5.5cm, trim={0.0cm 0.0cm 0.0cm 0.0cm}, clip]{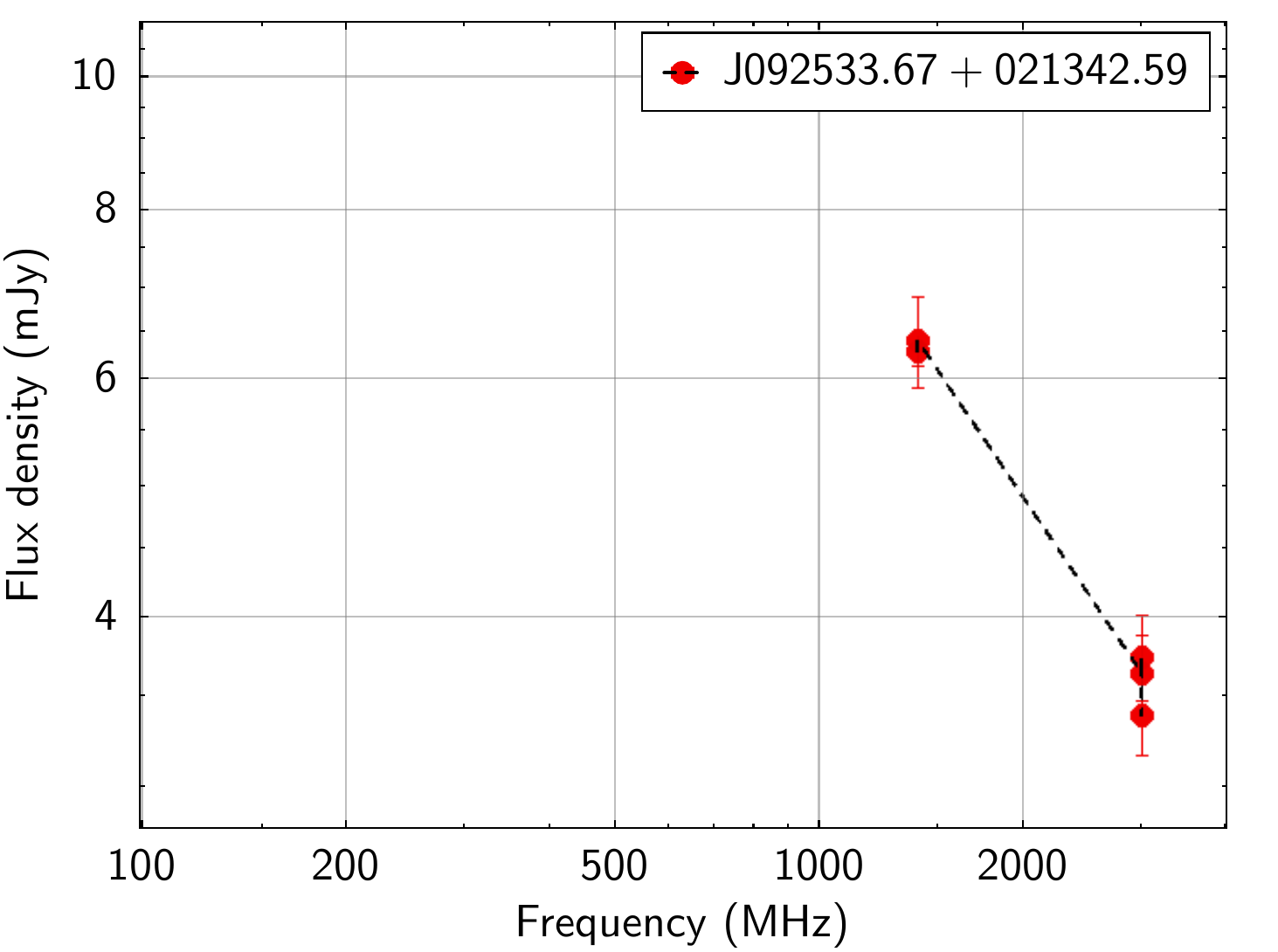}
\includegraphics[angle=0, width=5.5cm, trim={0.0cm 0.0cm 0.0cm 0.0cm}, clip]{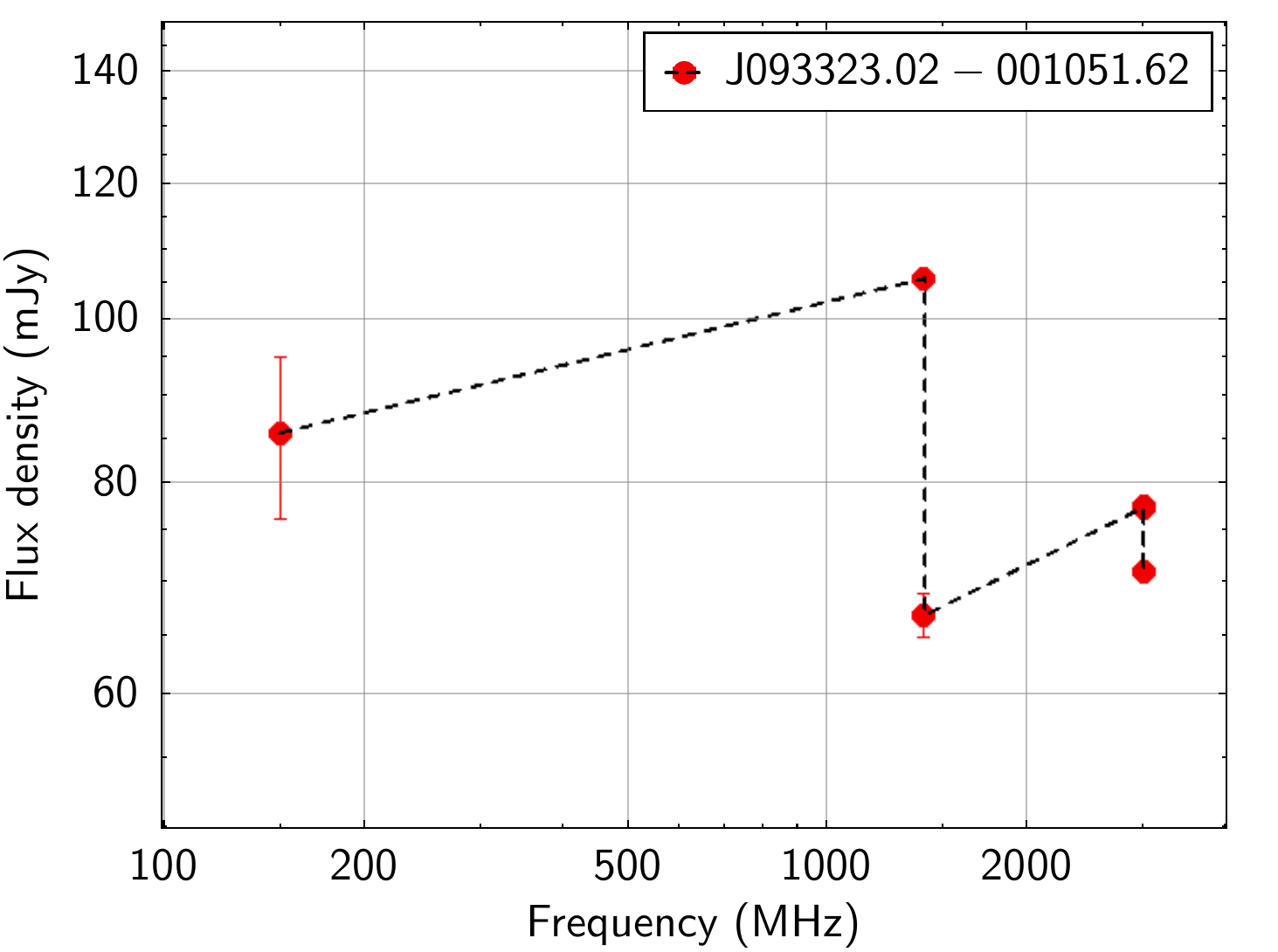}
\includegraphics[angle=0, width=5.5cm, trim={0.0cm 0.0cm 0.0cm 0.0cm}, clip]{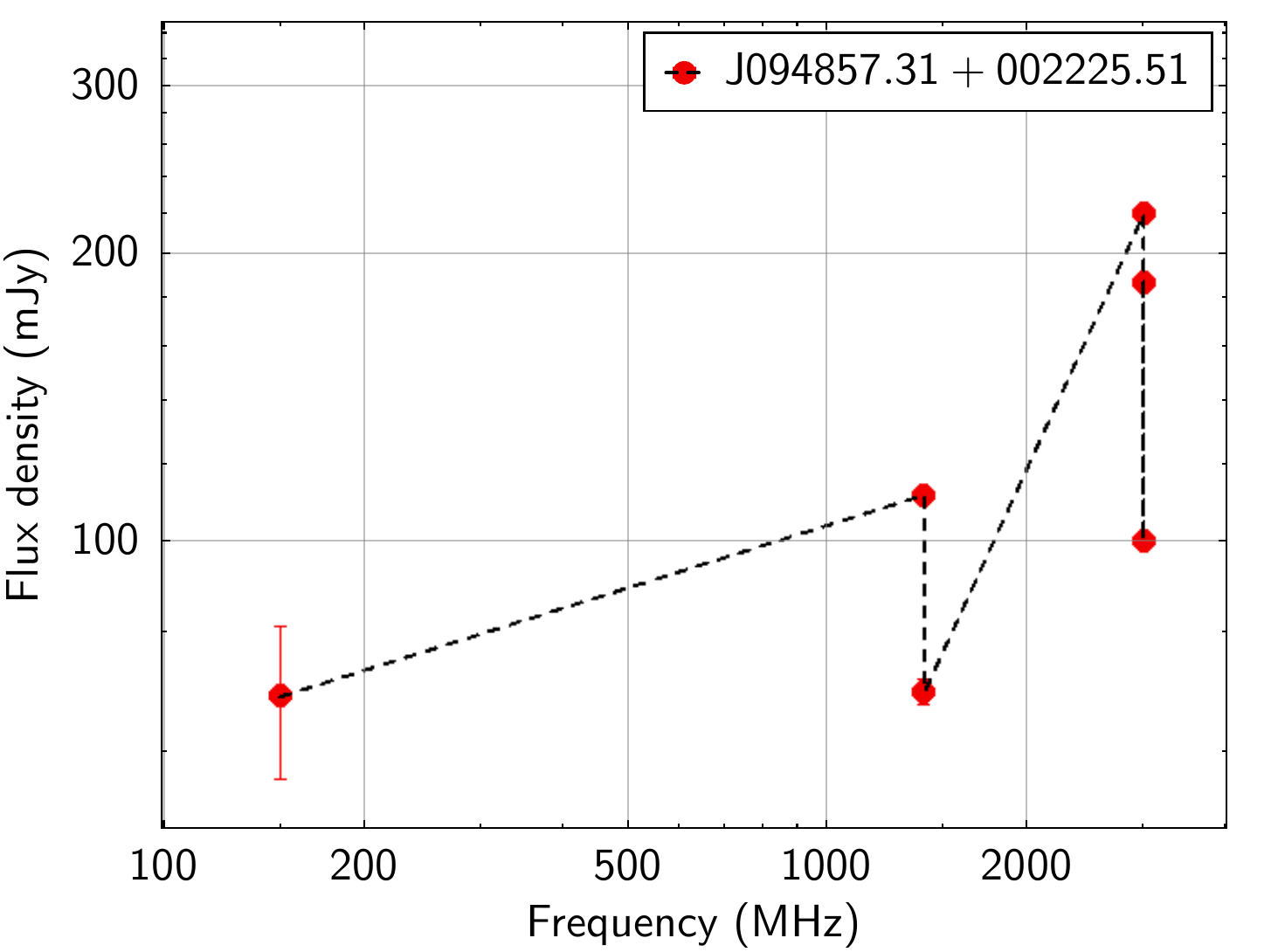}
\includegraphics[angle=0, width=5.5cm, trim={0.0cm 0.0cm 0.0cm 0.0cm}, clip]{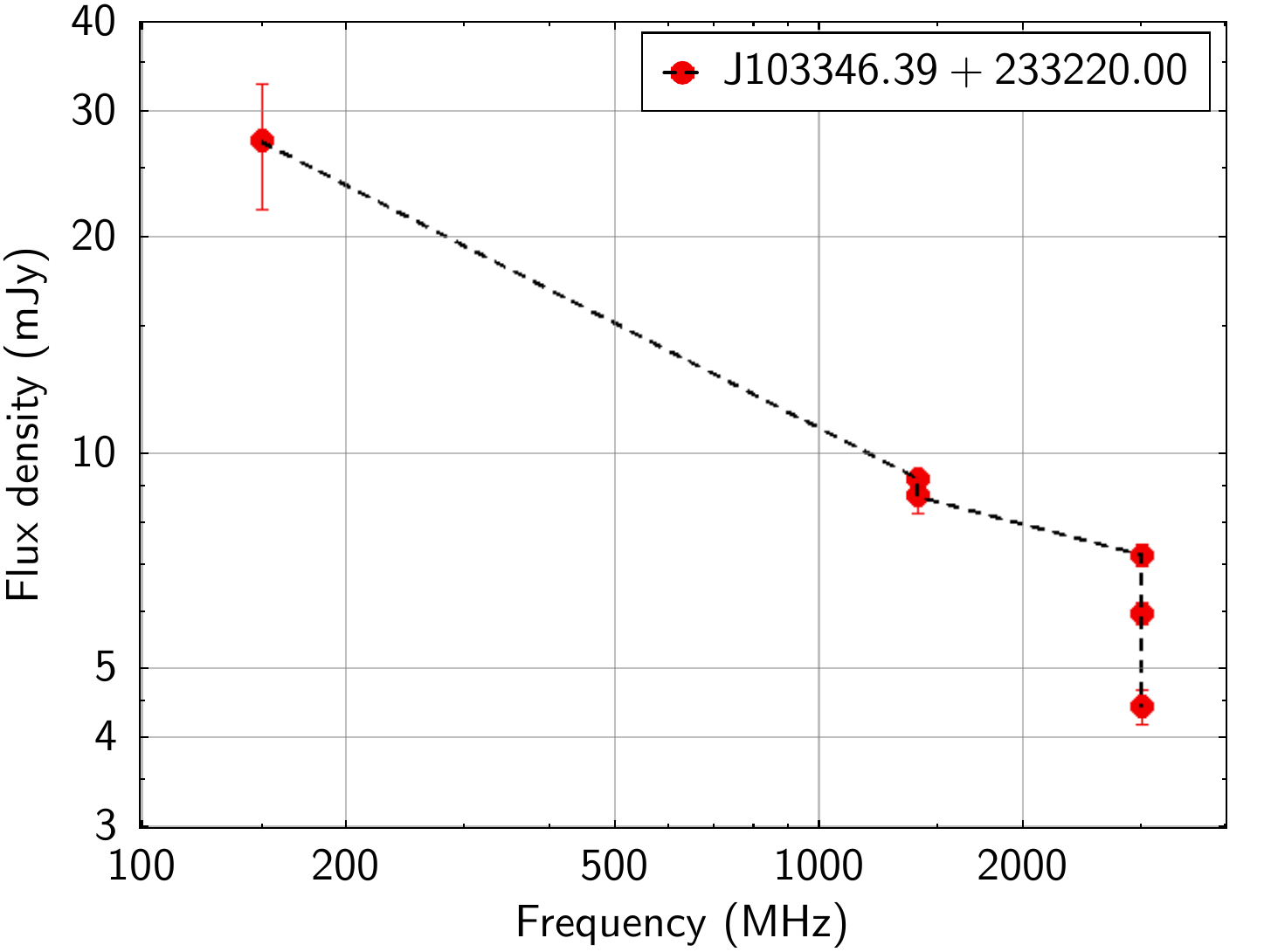}
\includegraphics[angle=0, width=5.5cm, trim={0.0cm 0.0cm 0.0cm 0.0cm}, clip]{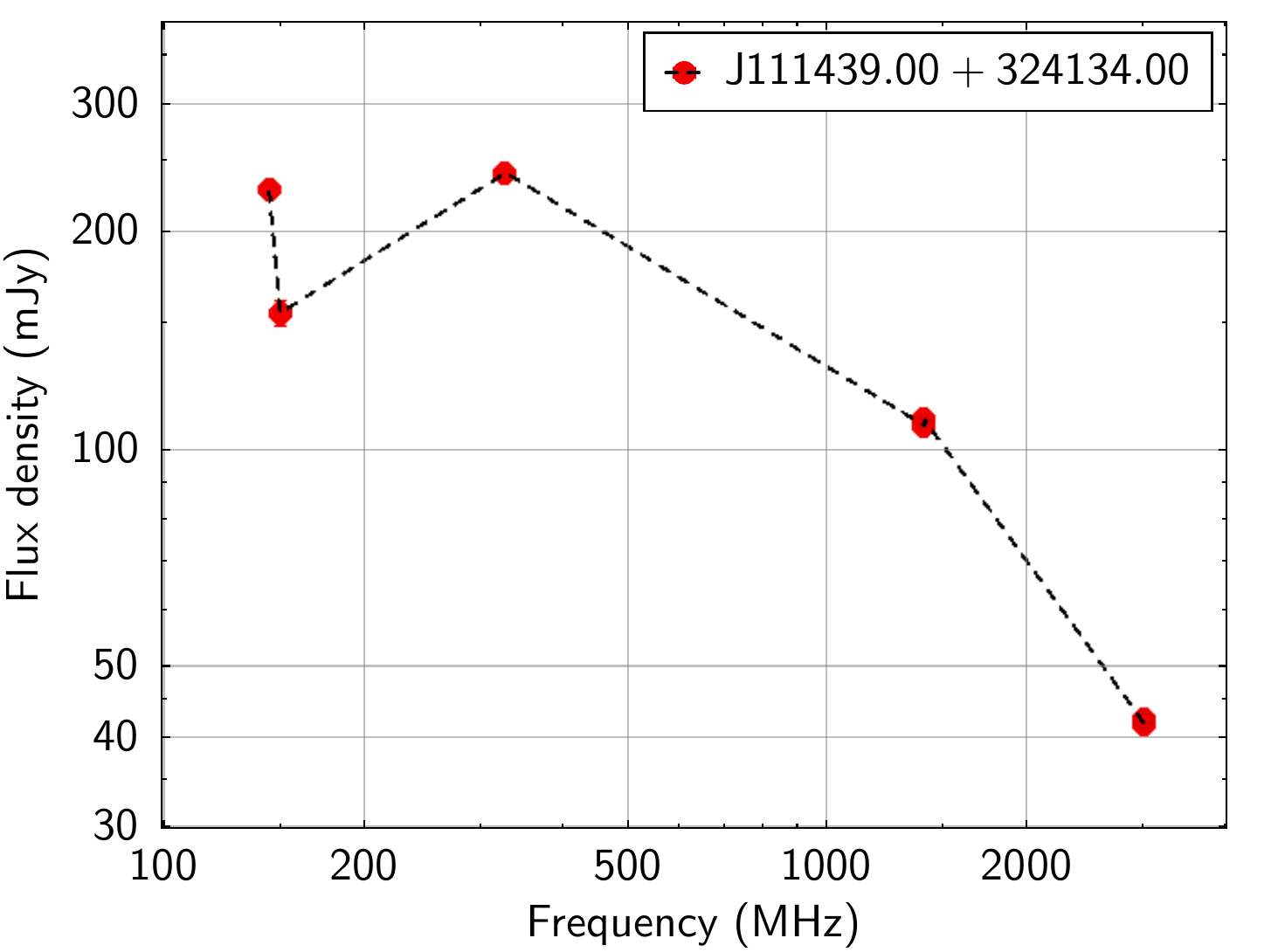}
\includegraphics[angle=0, width=5.5cm, trim={0.0cm 0.0cm 0.0cm 0.0cm}, clip]{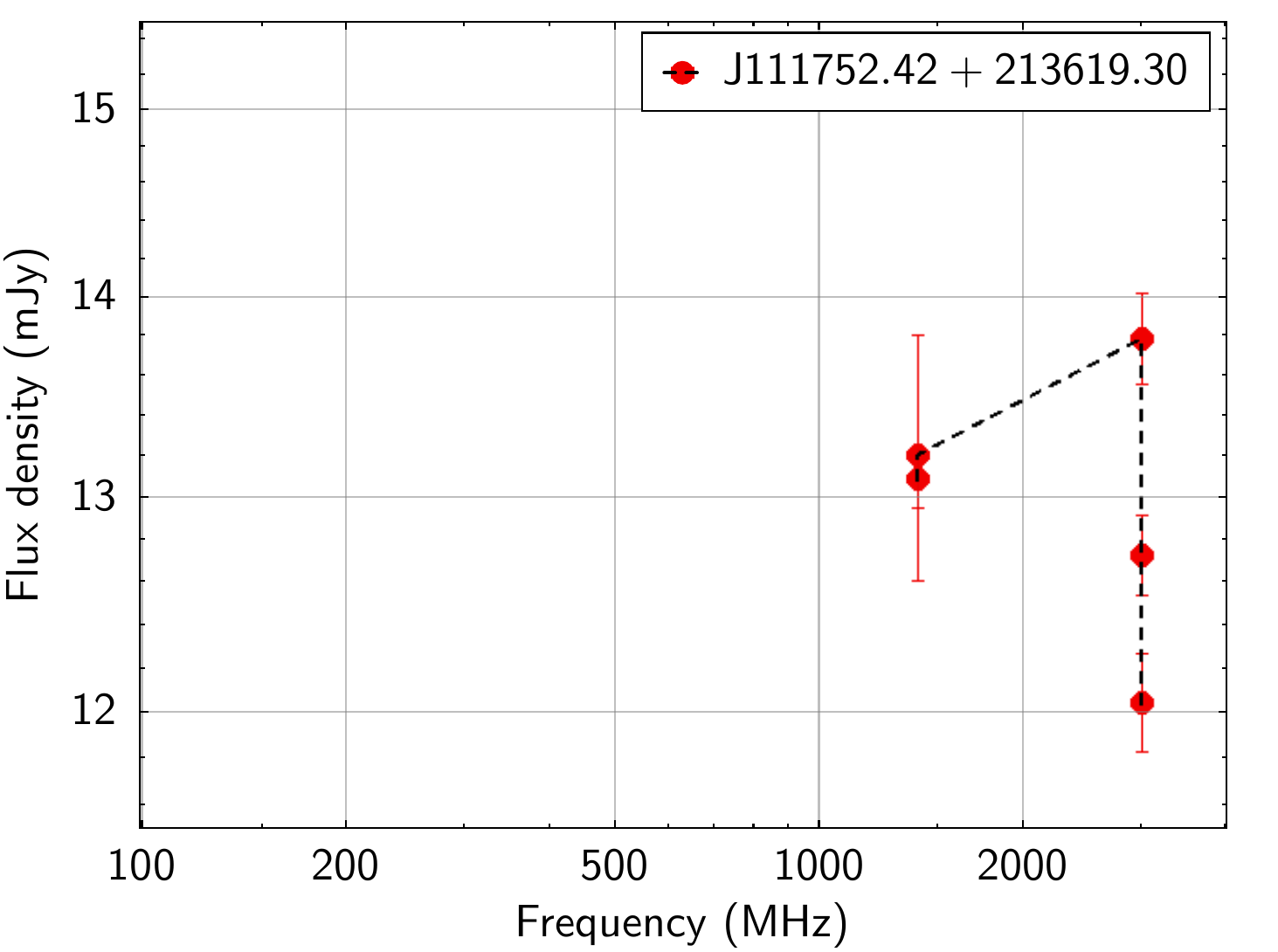}
\includegraphics[angle=0, width=5.5cm, trim={0.0cm 0.0cm 0.0cm 0.0cm}, clip]{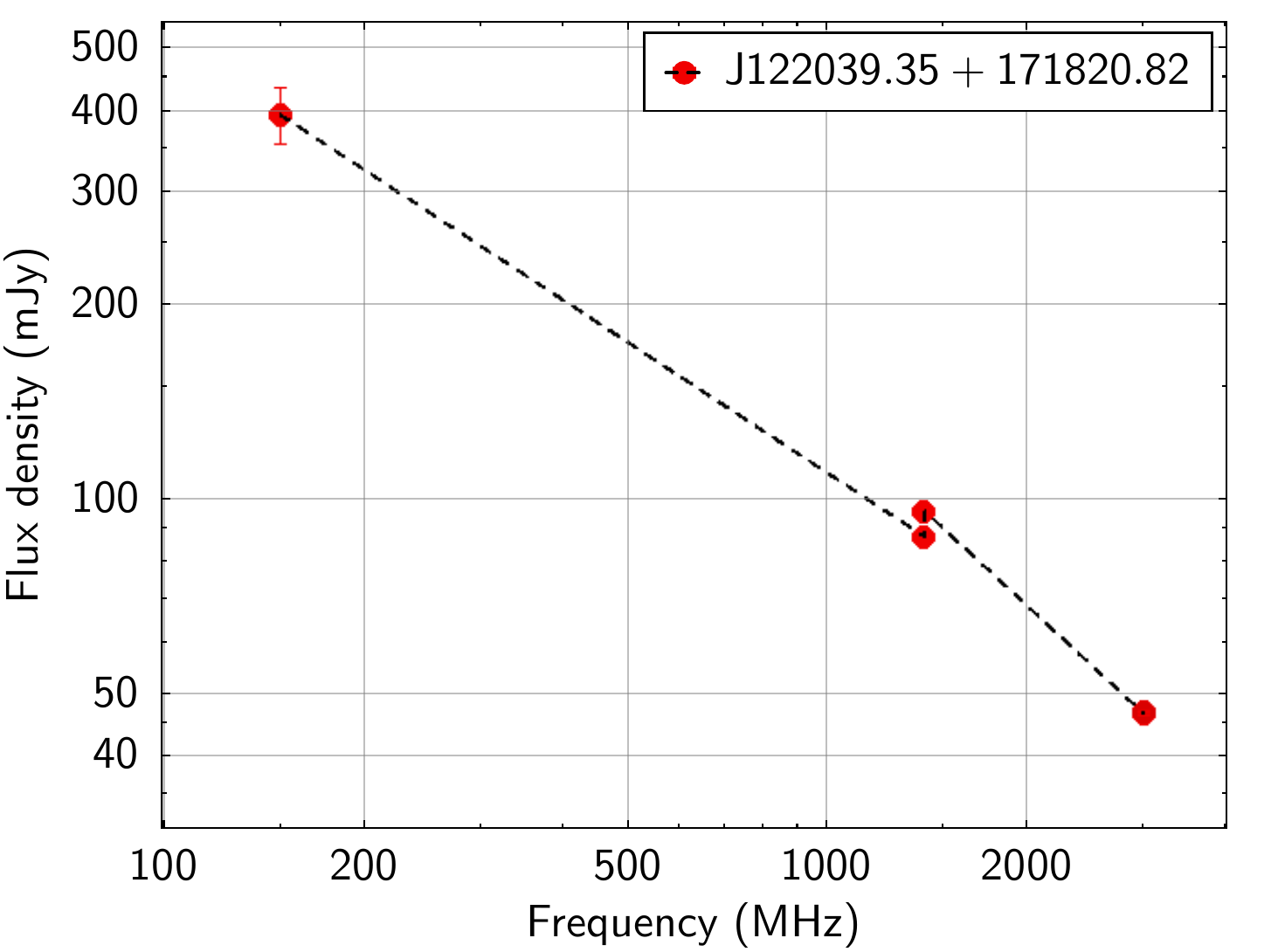}
\includegraphics[angle=0, width=5.5cm, trim={0.0cm 0.0cm 0.0cm 0.0cm}, clip]{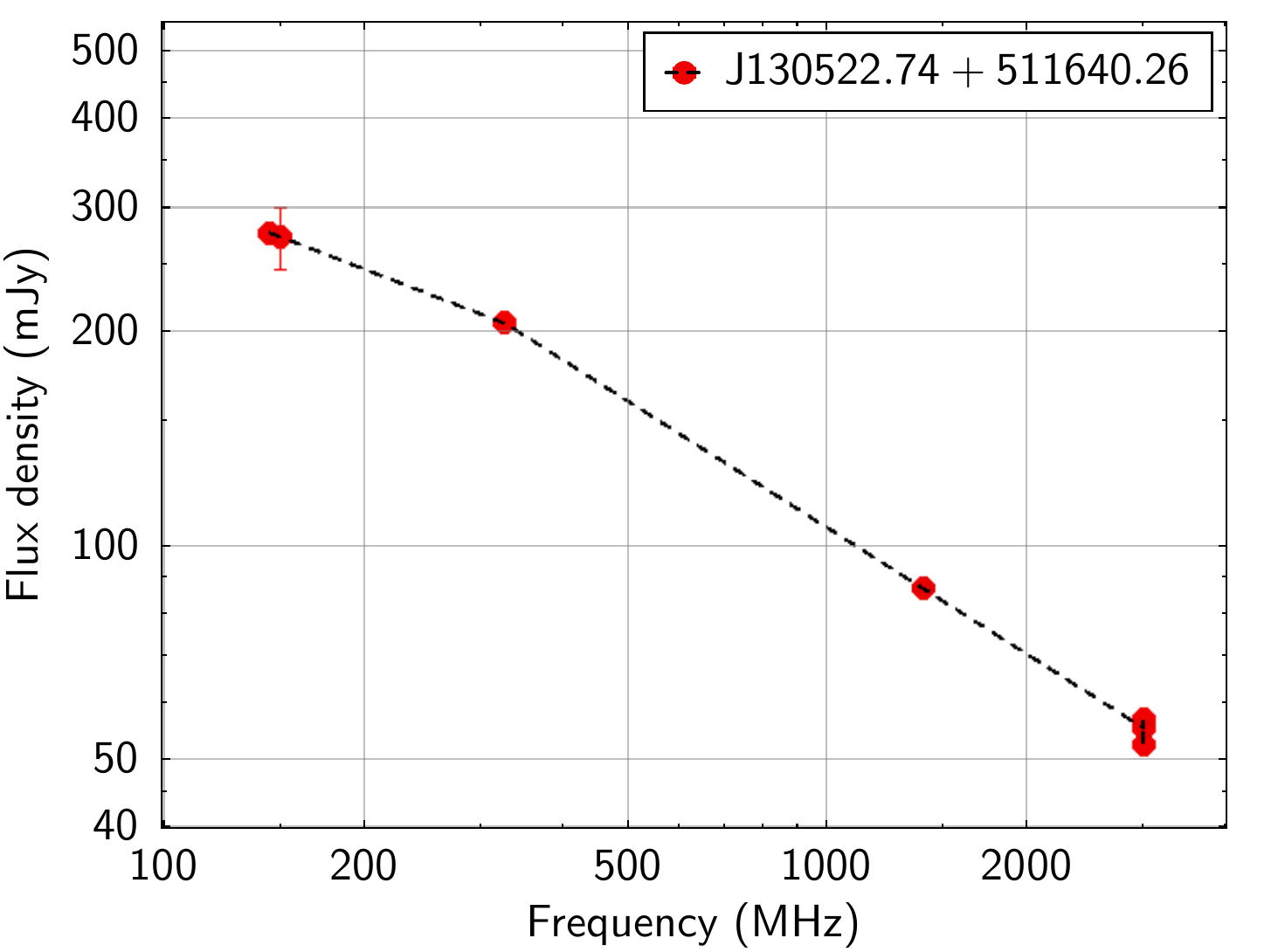}
\includegraphics[angle=0, width=5.5cm, trim={0.0cm 0.0cm 0.0cm 0.0cm}, clip]{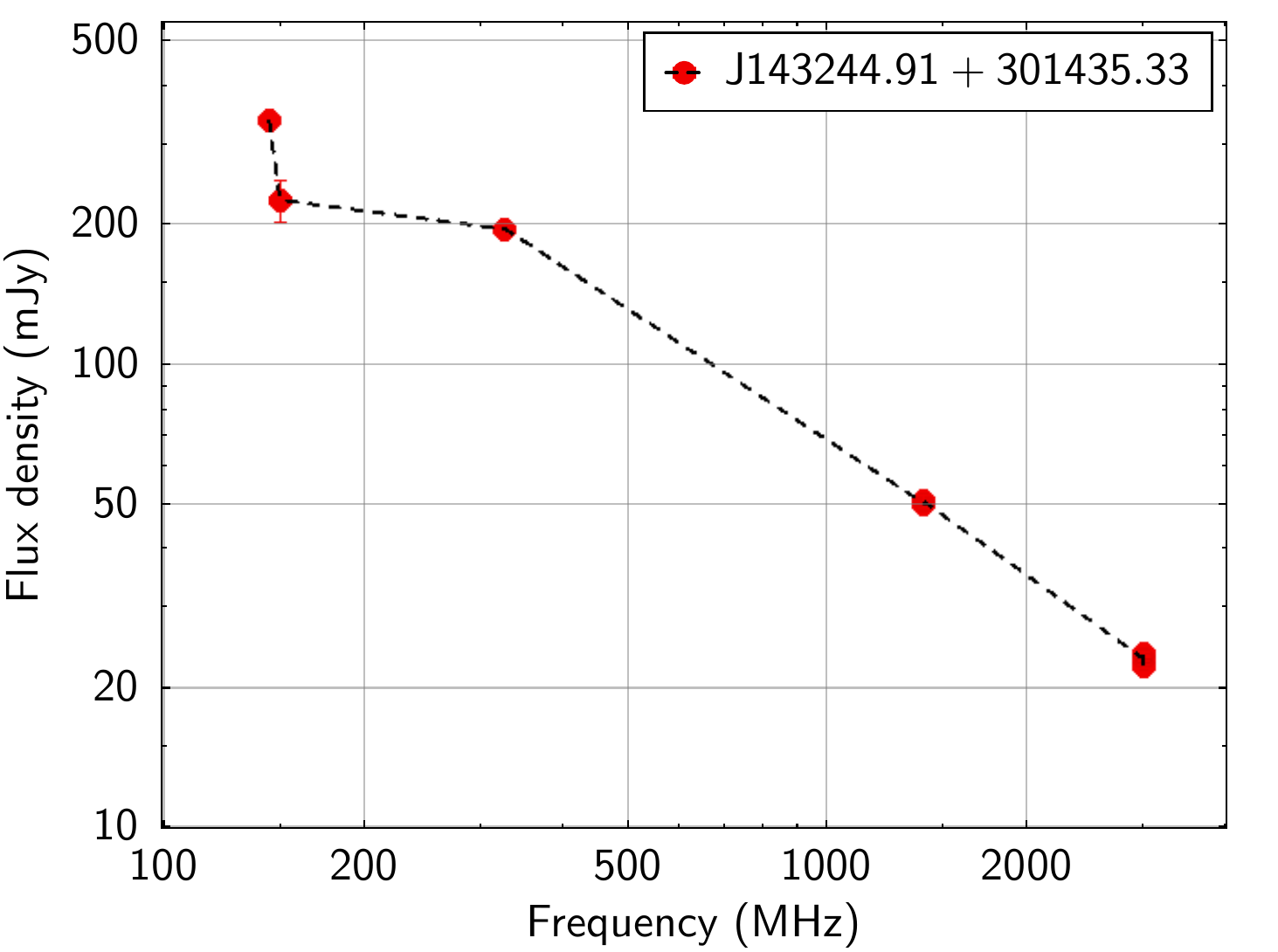}
\includegraphics[angle=0, width=5.5cm, trim={0.0cm 0.0cm 0.0cm 0.0cm}, clip]{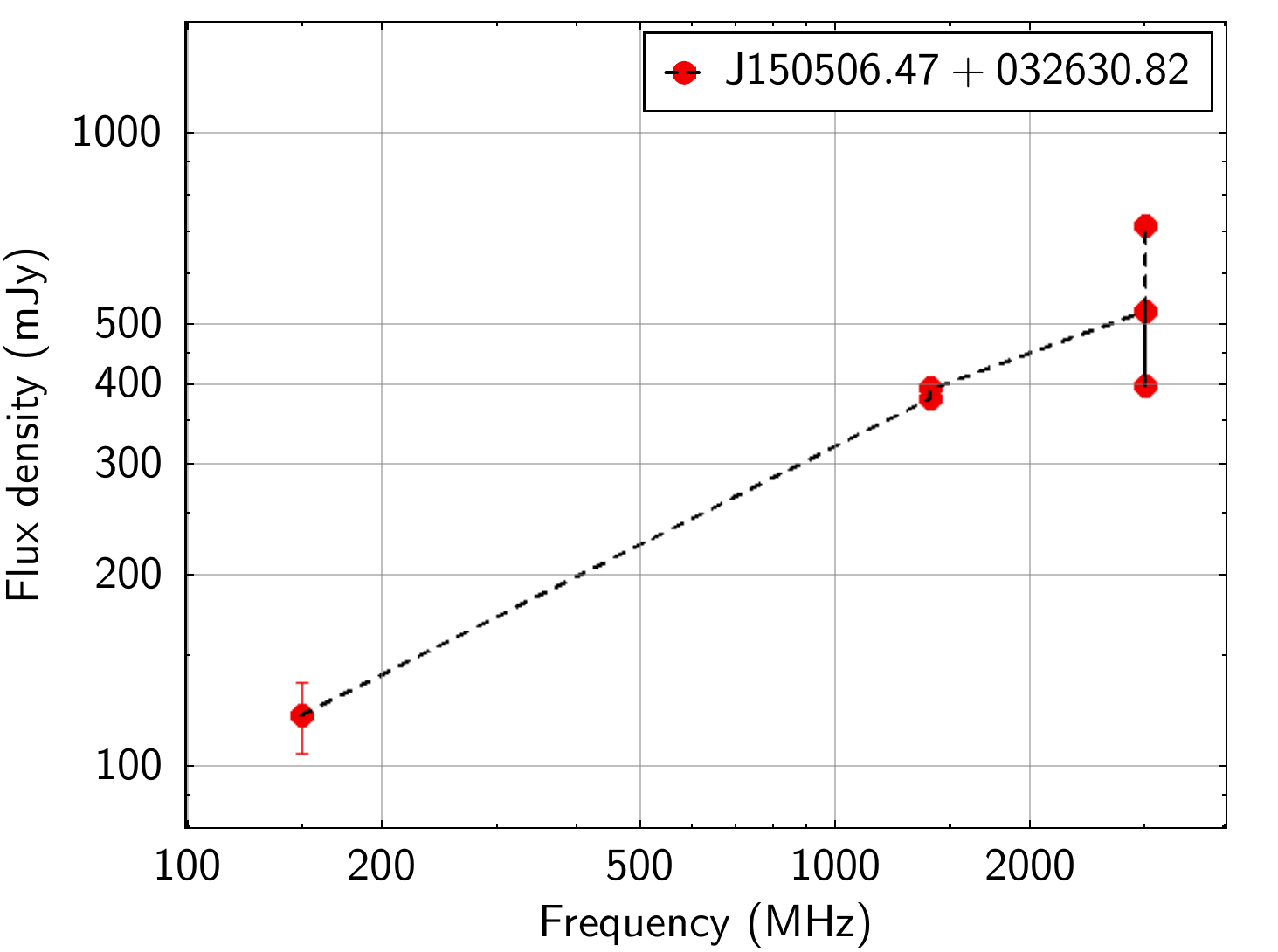}
\includegraphics[angle=0, width=5.5cm, trim={0.0cm 0.0cm 0.0cm 0.0cm}, clip]{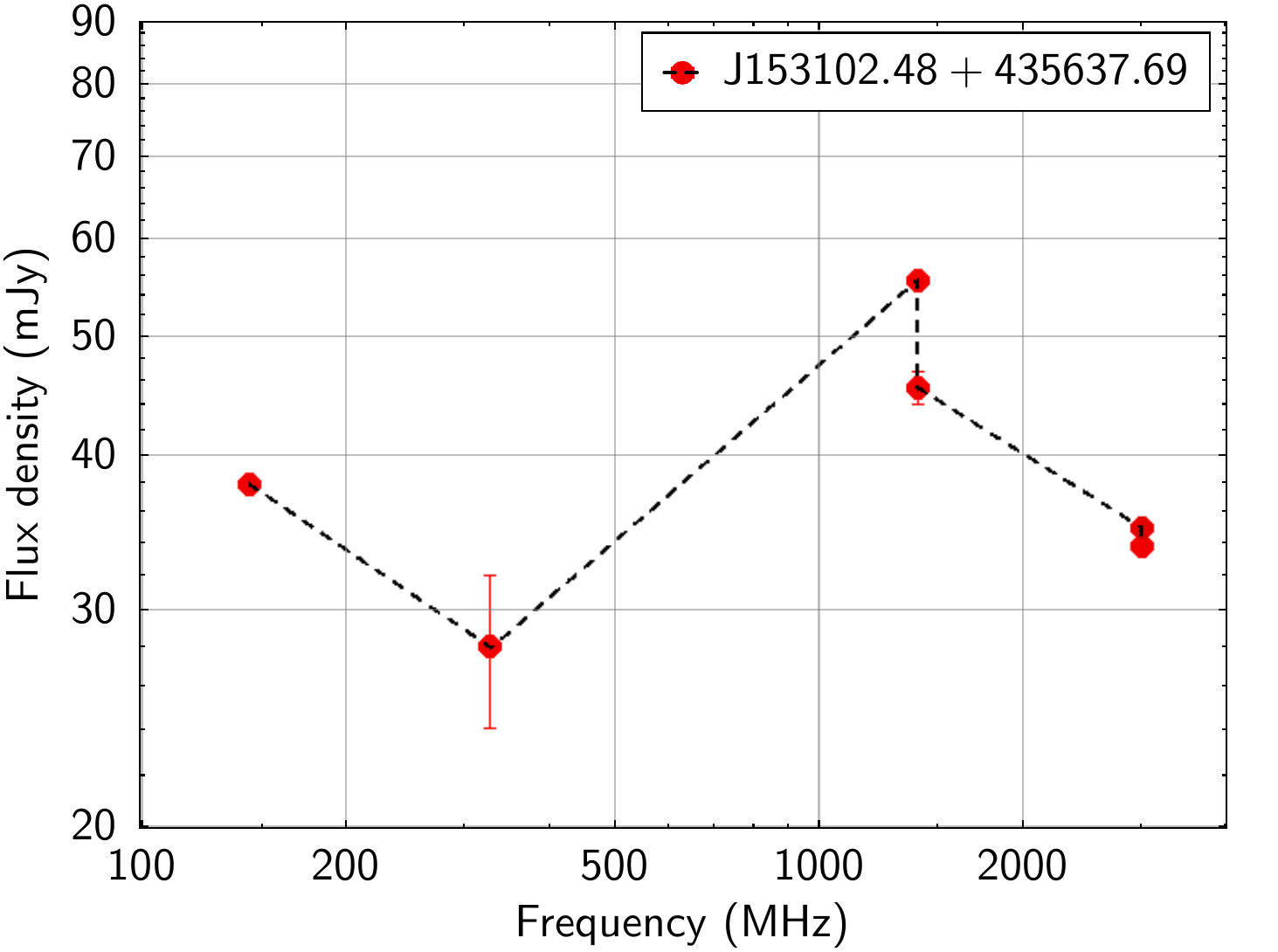}
\includegraphics[angle=0, width=5.5cm, trim={0.0cm 0.0cm 0.0cm 0.0cm}, clip]{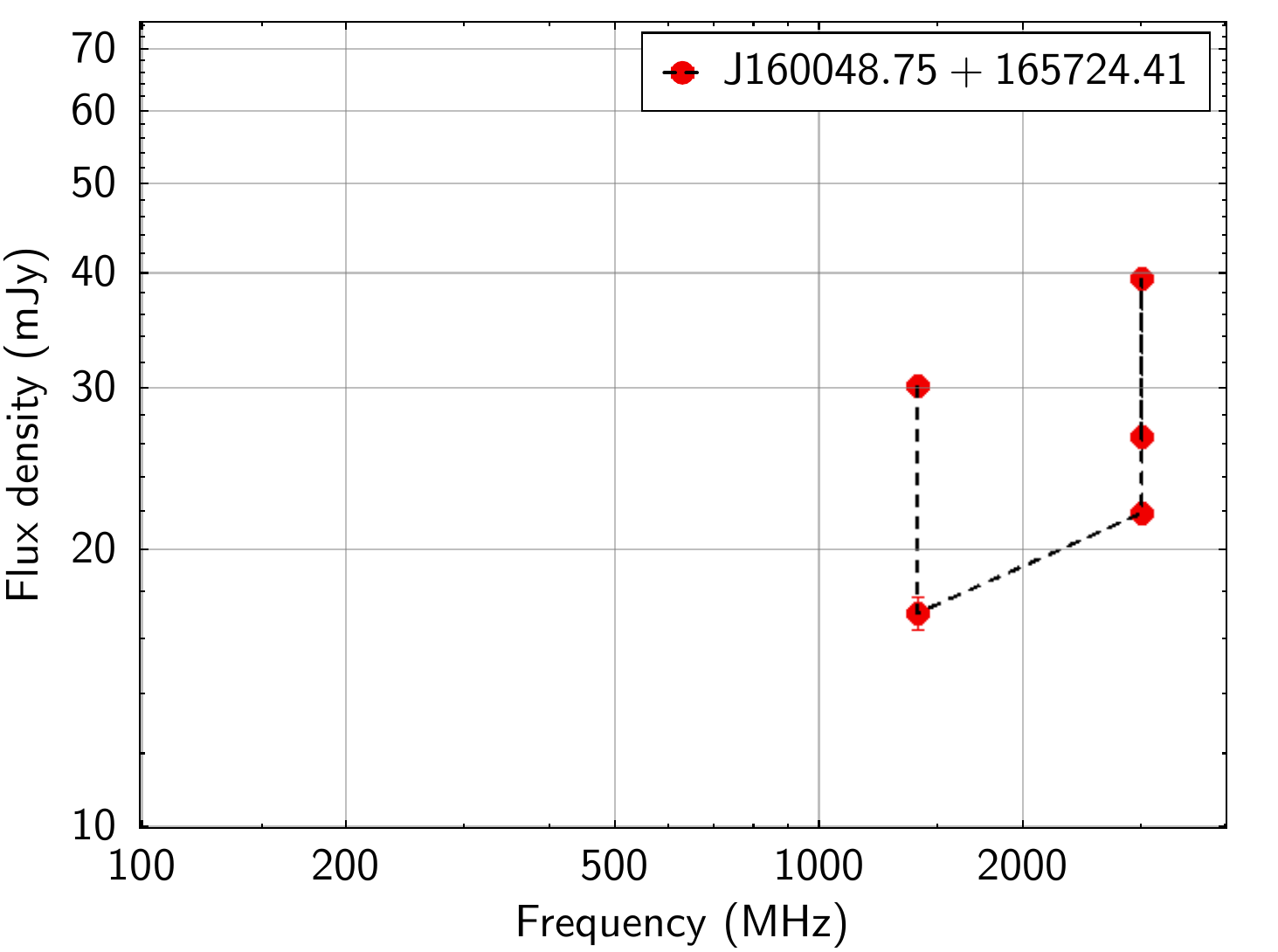}
\caption{The radio SEDs of our RL-NLS1s generated using non-simultaneous multi-frequency radio observations.
The spectral breaks seen in SEDs are attributed to radio variability.}
\label{fig:SEDs}
\end{figure*}

%
%

\bibliography{INOV}{}
\bibliographystyle{aasjournal}

\end{document}